\documentclass[11pt,leqno]{article}
\usepackage{graphicx}
\graphicspath{{./figures/}}

\usepackage{subfiles}

\usepackage{amsfonts}
\usepackage{amsmath}
\usepackage{amsthm}
\usepackage[export]{adjustbox}
\usepackage{booktabs}
\usepackage{varioref}
\usepackage{float}

\usepackage{color}
\usepackage{ amssymb }
\usepackage{hyperref}

\usepackage{natbib}

\newcommand\vep{{\varepsilon}}
\newcommand\bmu{\boldsymbol{\mu}}
\newcommand\bxi{\boldsymbol{\xi}}

\newcommand\bA{{\bf A}}
\newcommand\ba{{\bf a}}
\newcommand\bB{{\bf B}}
\newcommand\bd{{\bf d}}

\newcommand\mcH{{\mathcal H}}

\newcommand\bK{{\bf K}}
\newcommand\mbK{{\mathbb K}}
\newcommand\mcK{{\mathcal K}}

\newcommand\mcN{{\mathcal N}}

\newcommand\bS{{\bf S}}

\newcommand\bT{{\bf T}}

\newcommand\bx{{\bf x}}

\newcommand\bZ{{\bf Z}}

\newcommand{\blambda}{\boldsymbol{\lambda}}
\newcommand{\bzeta}{\boldsymbol{\zeta}}

\newcommand\convP{{\overset{P}{\to}}}

\usepackage{enumerate}

\newcommand\mcF{{\mathcal F}}
\newcommand\mbR{{\mathbb R}}

\DeclareMathOperator\E{E}
\DeclareMathOperator\diag{diag}
\DeclareMathOperator\Var{Var}
\DeclareMathOperator\Cov{Cov}

\DeclareMathOperator\logit{logit}

\newtheorem{lemma}{Lemma}
\newtheorem{theorem}{Theorem}

\newtheorem{assumption}{Assumption}

\theoremstyle{definition}

\newcounter{subassumption}[asu]

\makeatletter
\renewcommand{\p@subassumption}{\theasu}
\makeatother

\title{Highly Irregular Functional Generalized Linear Regression with Electronic Health Records}
\author{Justin Petrovich\thanks{\textit{Corresponding author.} Address: Alex G. McKenna School of Business, Economics, and Government, Saint Vincent College, 300 Fraser Purchase Rd., Latrobe, PA 15650, USA. Email: justin.petrovich@stvincent.edu }\\ Saint Vincent College, USA 
	\and Matthew Reimherr\\ Pennsylvania State University, USA
	\and Carrie Daymont\\ Penn State College of Medicine, USA}
\date{}

\begin{document}
\maketitle
\begin{abstract}

This work presents a new approach, called MISFIT, for fitting generalized functional linear regression models with sparsely and irregularly sampled data.  
Current methods do not allow for consistent estimation unless one assumes that the number of observed points per curve grows sufficiently quickly with the sample size. In contrast, MISFIT is based on a multiple imputation framework, which has the potential to produce consistent estimates without such an assumption.  Just as importantly, it propagates the uncertainty of not having completely observed curves, allowing for a more accurate assessment of the uncertainty of parameter estimates, something that most methods currently cannot accomplish.  This work is motivated by a longitudinal study on macrocephaly, or atypically large head size, in which electronic medical records allow for the collection of a great deal of data. However, the sampling is highly variable from child to child. Using MISFIT we are able to clearly demonstrate that the development of pathologic conditions related to macrocephaly is associated with both the overall head circumference of the children as well as the velocity of their head growth.
\end{abstract}


\section{Introduction}\label{sec:intro}
In recent years, Functional Data Analysis, FDA, has seen a rapid expansion into what \citet{wang2016functional} called \textit{next-generation functional data analysis}, as more complex applications and models are explored.  An ongoing challenge in this expansion is the handling of functional data that are either very irregularly sampled, sparsely sampled, or contain missing regions.  Classically, FDA is concerned with the statistical analysis of data where one or more variables of interest is a function.  If the functions are not completely or densely observed, then fitting certain FDA models becomes substantially more challenging. Extending FDA methods to handle such data has been a rich area of research \citep{shi1996analysis,brumback1998smoothing,james2000principal,rice2001nonparametric,yao2005functional}.    

One of the most common approaches for handling such data is to smooth or impute what is missing \citep{rice2001nonparametric}.  This imputation can be done on the curves themselves or on the scores in a Karhunen-Lo\'eve expansion (i.e. functional principal components) \citep{yao2005functional}.  This approach is especially attractive as it allows practitioners to subsequently draw upon a wide range of methods after imputation, either multivariate in the case of score level imputation or functional in the case of curve level imputation.  However, as it stands, nearly all methods presented in the literature carry out the imputation process while ignoring subsequent modeling that is to be done with the reconstructed curves or scores (a notable exception being Bayesian methods, e.g. \citet{thompson2008bayesian,kowal2019functional}). 
Such an approach can produce substantially biased estimates as well as produce unreliable standard errors and subsequent p-values.  Indeed, in many settings the resulting estimators need not even be consistent unless one assumes that the imputed curves converge to the truth asymptotically.  Such an assumption is mathematically convenient, but highlights a serious concern when handling sparse functional data.  
For these reasons, methods such as \citet{yao2005functionalR}, which does provide consistent estimation of functional linear models, avoid imputation and instead focus on a moment based estimation method combined with nonparameteric smoothing of covariances and cross-covariances.  However, this approach becomes difficult to extend to nonlinear models, and does not allow for the utilization of the vast literature on dense FDA methods.      

In this work we present a new approach that provides a framework for developing consistent estimation techniques for both linear and nonlinear models, while also presenting a more complete bridge between sparse and dense functional data.  In particular, we attack the problem by pivoting our perspective to that of a missing data problem.  In the context of the literature on missing data, the goal is to impute the missing data in a way that preserves the performance of subsequent statistical modeling.  We show how combining ideas from PACE \citep{yao2005functional} and multiple imputation \citep{rubin1996multiple,schafer1999multiple,rubin2004multiple,royston2004multiple} results in an approach that alleviates many of the discussed issues, while also remaining quite broadly applicable.  Given that our foundation is built upon multiple imputation, we call our new method {\it Multiple Imputation of Sparsely-sampled Functions at Irregular Times} (MISFIT). As a by-product of this work, in Sections \ref{sec:logistic} and \ref{sec:cont} we also present results connecting regression, imputation, and the equivalence of probability measures, which are of independent interest.

\subsection{Macrocephaly}
Clinical data have long been analyzed using longitudinal data methods, which enable one to account for the correlation between different measurements on the same subject. However, if one also assumes that these repeated measurements constitute realizations of a smooth curve or data-generating process, modeling the data as functional data can be advantageous both in terms of flexibility and statistical power \citep{he2011functional,szczesniak2016longitudinal,craig2017infant,goldsmith2017variable}. Since clinical visits may occur both infrequently and irregularly, their analysis poses a challenge to the current smoothing/imputing methods discussed in section \ref{sec:intro}. To illustrate the effectiveness of our approach at addressing these challenges, we apply it to one such clinical data set in order to predict the presence or absence of pathologic conditions related to macrocephaly.

Head circumference is routinely measured in children between birth and two years of age, primarily for the purpose of detecting pathologic conditions that cause atypically large head size (macrocephaly) and atypically small head size (microcephaly). Particularly for conditions causing macrocephaly--which include hydrocephalus, brain tumors, and chronic subdural bleeding (often caused by abusive head trauma)--delayed identification and treatment may lead to poorer long-term outcomes. Most children with a large head are healthy, and distinguishing healthy children with a large head from children with pathologic conditions causing head enlargement is challenging. Expert opinion-based methods for evaluating head size that are used by clinicians have been shown to discriminate poorly between children with and without pathology \citep{daymont2012test,wright2015head}. Delineating features of a child's head circumference trajectory that are predictive of pathology may improve identification of children at high risk. For example, automated evaluations of head circumference trajectories could be incorporated into electronic health record-based tools to signal to clinicians that further evaluation may be warranted. 

Because pathology associated with macrocephaly is rare, research in this area often requires use of existing clinical data from electronic health records, rather than prospective research with high-quality measurements performed at defined intervals. Based on the typical schedule for preventive health care visits \citep{workgroup20172017}, it is rare for a child in the U.S. to have more than 10 head circumference measurements, and many children will have fewer. There is significant variability in the timing of appointments, and clinical measurements are affected by errors of varying type and degree \citep{daymont2017automated}. Similar challenges apply to other growth measurements, such as weight and length, as well as other types of clinical measurements. The ability to characterize trajectories of sparse irregular data has potential applicability to many clinical questions. 

The growth data evaluated in this paper were extracted from the electronic health record of a large primary care network\citep{daymont2010head}. Manual chart review was used to identify children, aged between 3 days and 3 years, with pathologic conditions that are known to be associated with macrocephaly, as described in detail previously\citep{daymont2012test}. Of 74,428 children, 85 with pathologic conditions were identified.

\begin{figure}[h]
  \centering
  \includegraphics[scale = .4]{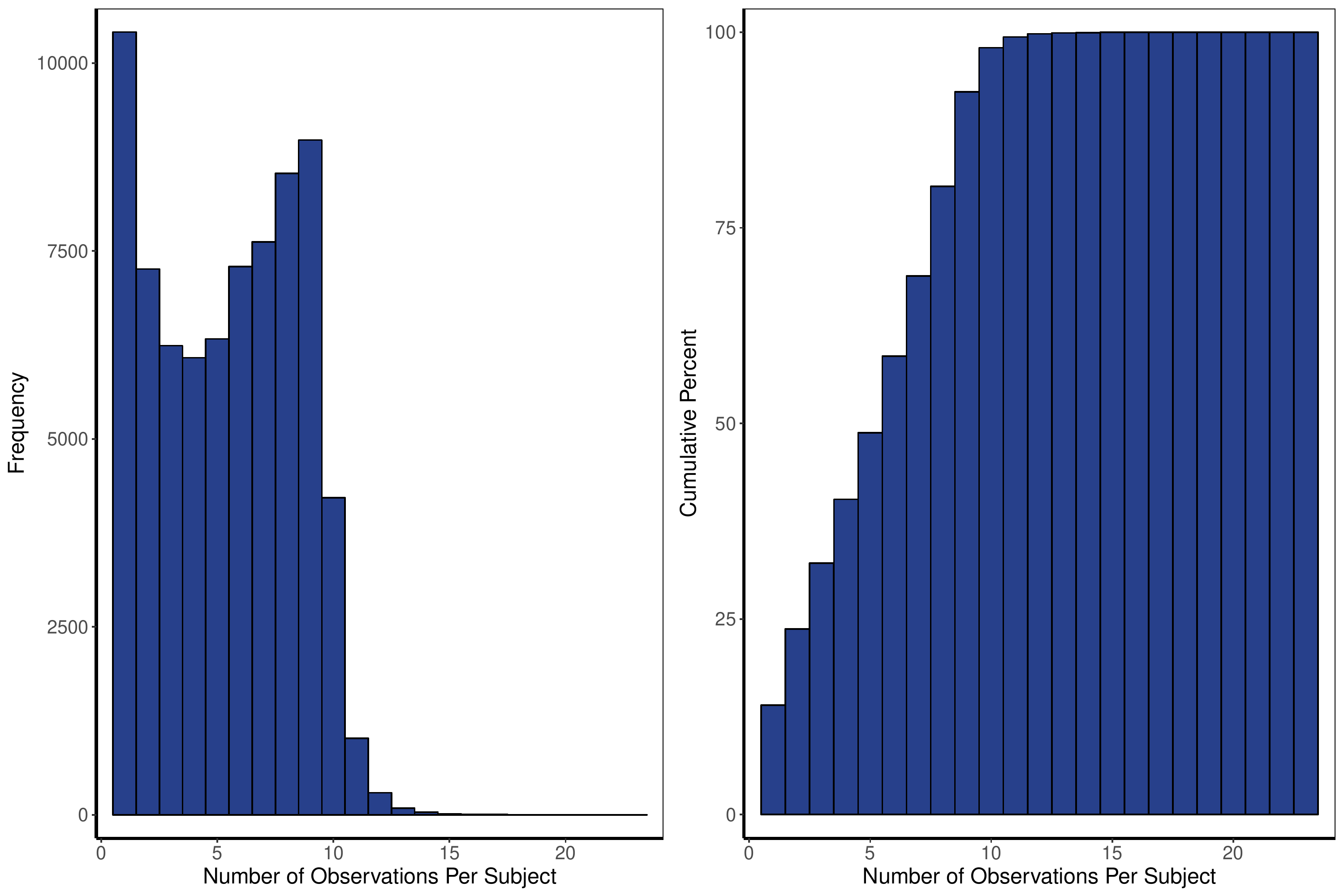}
  \caption{Left panel: histogram of the number of observations per subject, ranging from 1 to 23. Right panel: histogram of the cumulative percentage of observations per subject.}
  \label{fig:1}
\end{figure}

Though the term sparsity is somewhat subjective in the context of functional/longitudinal data, many of the subjects in the present data set have just a single measurement, while the maximum number of measurements for any subject is 23. In the left panel of figure \ref{fig:1} is a histogram of the number of observations per subject in the data, clearly illustrating that the modal number of measurements is 1, while relatively few children had more than 10 clinical visits and almost none had more than 15. The right panel of figure \ref{fig:1} is a cumulative histogram of the relative frequency of number of observations, allowing us to determine the proportion of subjects who had no more than a given number of observations. Specifically, about 98\% of subjects received 10 or fewer measurements, 49\% were measured at most 5 times, 24\% were observed no more than twice, and 14\% attended a single clinical visit for measurement.

\begin{figure}[h]
  \centering
  \includegraphics[scale = .4]{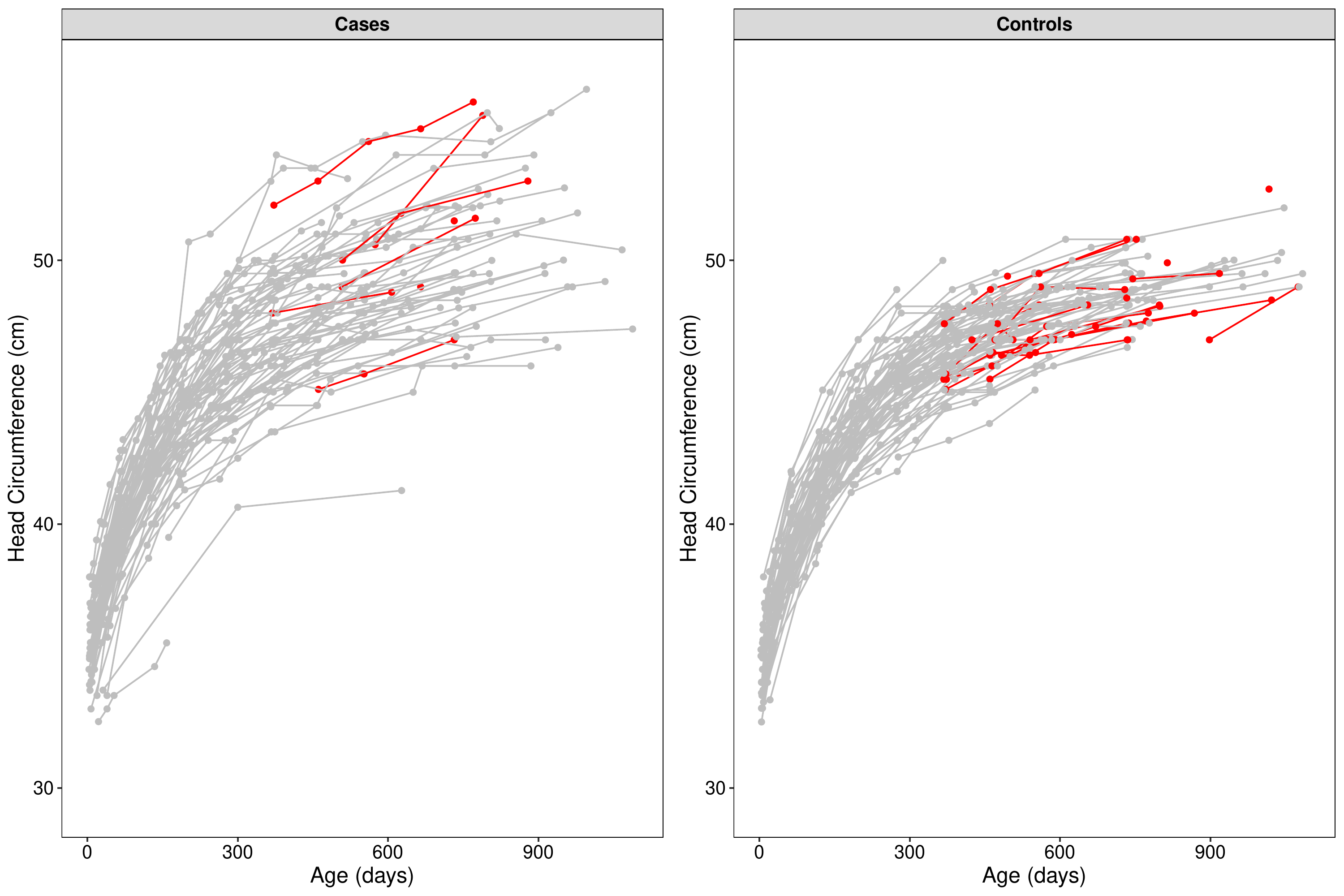}
  \caption{Spaghetti plots of head circumference trajectories for all 85 cases (left panel) and 100 randomly selected controls (right panel) from the data. Red lines highlight subjects whose first visit occurred when they were at least 1 year old.}
  \label{fig:2}
\end{figure}

In addition to the dearth of observations for many of the subjects in the data set, subjects were not observed with any uniformity or regularity. Figure \ref{fig:2} illustrates this and provides a glimpse of the data. While nearly half of all subjects in the data set were observed once by the time they were a month old, notice that for both the cases and the controls, several subjects were not observed until they were at least 1 year old (represented by the red lines). Furthermore, we see again several subjects with a single observation (a single dot with no attached line), and can clearly tell that visits are not guaranteed to occur at the same ages for all subjects. Having identified the head circumference trajectories as both sparse and irregular, we proceed to introduce MISFIT as an approach that accounts for these conditions in a functional regression framework before revisiting these data in Section \ref{sec:app}.

\subsection{Previous Work}
There has been extensive work on functional data analysis with non-densely sampled data. Maybe the most widely used perspective is that of PACE \citep{yao2005functional}.  In the context of missing data methods, PACE is a mean imputation method, though it is traditionally carried out on the principal component scores.  PACE can be carried out using an extensive software package in {\tt Matlab}, as well as a relatively recent port to {\tt R}, which has facilitated its implementation in a variety of applications \citep{chiou2014functional,liu2014sparse,peng2008distance}.  

While PACE is based on using local polynomial smoothing to estimate the unknown parameters, a similar approach based on splines can be found in the {\tt refund} package in {\tt R} \citep{staniswalis1998nonparametric,di2009multilevel,goldsmith2013corrected}.  This approach, like many of the methods in {\tt refund}, uses a more explicit mixed effects framework to impute the scores/curves.  In both cases, the primary idea is to borrow information across units to help with the imputation process.  However, these methods carry out the imputation without consideration of subsequent statistical analyses.  We will show that such an approach can lead to biased and even inconsistent estimators.  

It is worth noting that each of the above works has addressed the problem through the lens of a sparse functional data issue, but not from a missing data perspective. In \citet{rubin2004multiple}, three different missingness paradigms are delineated: 1) Missing Completely at Random (MCAR), in which the missing value patterns are independent of all data 2) Missing at Random (MAR), in which the missing value patterns depend only on the observed data and are conditionally independent of the unobserved data 3) Missing Not at Random (MNAR), in which the missing data patterns depend on the observed and unobserved data. While there is no explicit treatment of a missing data mechanism in any of the aforementioned papers, it is implicitly assumed that the researcher is in one of the two former paradigms, either MCAR or MAR. We make the same implicit assumption, without formally defining the missing data mechanism in our procedure.

Several recent papers have built upon these ideas and adopted a missing data perspective to address various forms of nonresponse or sparsity in functional regression models. For example, \citet{he2011functional} used a Bayesian approach to handle missingness in the response variable of a longitudinal model. They employ a Gibbs sampler to draw model parameters as well as imputed values from their posterior distribution. In the spirit of multiple imputation, they propose multiple draws of imputed values upon convergence of the chain, or to simply use multiple independent chains and form one completed data set from each of the converged chains. Unlike our approach however, they focus on the functional mixed model and address only sparsity in the functional response variable.

Focusing on a scalar-on-function linear model, \citet{crambes2017regression} use an FPC-based estimator of the coefficient function to impute missing values of the scalar response by incorporating the missing data mechanism (assuming MAR) into the usual predictor for the response. In the same setting, \citet{ferraty2013mean} study estimation of the mean of a response variable, providing one method based on averaging predicted values and another based on propensity scores. Again, these works are distinct from ours since they focus on missingness in the response variable only. Furthermore, the former considers only a linear scalar-on-function regression model (while we consider more general generalized linear regression models), and the latter is not concerned with regression, but rather (unconditional) mean estimation of the response variable.

More akin to our set-up, \citet{preda2010nipals} use the NIPALS (nonlinear iterative partial least squares) algorithm to impute missing data in the functional covariates of a scalar-on-function linear model, where they assume that the missingness mechanism follows a two-state, continuous time markov process with exponential holding times. 
While the PLS-based NIPALS algorithm provides a nice alternative to regression on the FPCs, it still neglects to propagate uncertainty due to incomplete functional observations. In addition, these authors also focused only on the linear scalar-on-function regression model, leaving extensions to nonlinear models unexplored. As our motivating example requires a logistic regression set-up, none of these methods are immediately applicable.

\subsection{Organization}
The remainder of the paper is organized as follows.  We present our framework in Section \ref{sec:meth} for both linear regression and generalized linear regression with a natural exponential family.  
For glms, we consider both the case of categorical and continuous outcomes.
As a by-product, we present some interesting results concerning the relationship between functional regression and the equivalence of Gaussian measures.  In Section \ref{sec:sims} we present a numerical study that highlights the limitations of previous approaches and demonstrates how MISFIT fixes many of these issues.  In Section \ref{sec:app} we apply MISFIT to the evaluation of head circumference trajectories.  We show how our approach sheds insight into the relationship between head circumference growth and the presence of a pathology related to macrocephaly.  We also show how other approaches dramatically underestimate the uncertainty in this application. Section \ref{sec:theory} provides an initial theory in the linear case, highlighting how MISFIT produces consistent estimates as long as the imputation model can be consistently estimated.  Finally, in Section \ref{sec:con} we finish with concluding remarks and future research directions, especially as they pertain to more complicated models and deeper statistical theory.

\section{Methods}\label{sec:meth}
In this section, we detail imputation procedures for both a linear and generalized linear scalar-on-function regression model. This involves specifying an imputation model, and the parameters of the conditional distribution from which imputed values will be drawn.  In \ref{sub:setup} we provide necessary notation. Then, in \ref{sub:lm} we outline our procedure specifically for a linear regression model, motivating our MISFIT approach by more explicitly pointing to some difficulties with the PACE approach. Subsection \ref{sec:logistic} extends our method to categorical outcomes, where we develop an imputation model that is compatible with the scalar-on-function logistic regression model and connect this to the equivalence of Gaussian measures.  In Subsection \ref{sec:cont} we consider more general continuous outcomes.  Throughout, we assume a canonical link function, which turns out to be more than mere technicality, as it is required to assure a Gaussian imputation framework is compatible with GLM.  Finally, in \ref{sub:comp} we outline our computational strategy to implementing these methods.

\subsection{Setup and Notation}\label{sub:setup}
We denote the underlying functional covariates as $\{X_{i}(t): t \in [0,1]; 1 \leq i \leq N\}$, where $t$ denotes the argument of the functions, usually time, and $i$ denotes the subject or unit.  However, we assume that these curves are only observed at times $t_{ij}$ for $j = 1,\dots, m_i$, and with error:
\[
x_{ij} = X_i(t_{ij}) + \delta_{ij}.
\]
We let $\bx_i = (x_{i1}, \dots, x_{im_i})^\top$ denote the vector of observed values on the function $X_i$. 
Explicit distributional assumptions will be made later on.  

We assume that we have an outcome, $Y_i$, that is related to $X_i$ via a link function $g$:
\[
\E[Y_i | X_i] = g^{-1}(\eta_i) \qquad \eta_i = \alpha + \int X_i(t) \beta(t) \ dt.
\]
Throughout, when integration is written without limits, it is understood to be over the entire domain, in this case $[0,1]$.  
The goal of this work is to develop tools for consistently estimating $\alpha$ and $\beta(t)$.  As we will see, using PACE to first produce scores or curves and then fit the corresponding model will not, in general, result in a consistent estimate unless one can guarantee that the smoothed/imputed curve actually converges to the truth as the sample size grows.

\subsection{Linear Models} \label{sub:lm}
     In this section we assume that the outcome, $Y_i$, is continuous and related linearly to a functional predictor $X_i(t)$:
\begin{align}
Y_i =  \alpha  + \int \beta(t) X_i(t) + \vep_i. \label{e:sonf}
\end{align}
We assume that the $X_i(t) \in L^2[0,1]$ are i.i.d. Gaussian processes with mean $0$ and covariance function $C_X(t,s)$.  The $X_i(t)$ can then be expressed using the well-known Karhunen-Lo\`{e}ve expansion:
\[
X_i(t) = \mu_X(t) + \sum_{j=1}^\infty \xi_{ij} v_j(t).
\]
The $v_j(t)$ are the eigenfunctions of $C_X$ with corresponding eigenvalues $\lambda_1 \geq \lambda_2 \geq \dots \geq 0$.  The scores are given by $\xi_{ij} = \langle X_i - \mu_X, v_j\rangle$, which are independent (across both $i$ and $j$) mean-0 normal random variables with variance $\lambda_j$.  
We also assume the errors are iid $\vep_i \sim N(0,\sigma_\vep^2)$, $\delta_{ij} \sim N(0,\sigma^2_\delta)$ and mutually independent.   

The challenge in estimating $\beta(t)$ is that the $X_i(t)$ are not densely observed, and what is observed is observed with error. Thus directly smoothing the $x_{ij}$ to plug into a dense estimation framework can result in substantial bias. PACE solves this problem by estimating the unknown parameters via pooled nonparametric smoothing, and then using them to form Best Linear Unbiased Predictors (BLUPs) of the curves/scores.  One can then plug those BLUPs into a dense estimation framework.  This approach is sensible and tends to work much better than direct smoothing of the $x_{ij}$, however it still suffers from at least two major problems that our procedure addresses.  First, the imputation in PACE is done without using the outcome $Y_i$ or any consideration of the subsequent models that will be fit.  This can result in a  biased estimate of $\beta(t)$, as we will soon show.  The second problem is that the uncertainty of the imputation is not incorporated into PACE-based procedures when forming confidence or prediction intervals, or p-values.  While there are established ways to eliminate the bias problem in the linear case, the second problem is still relatively under explored.

To better understand the bias problem, consider a naive FPCA-based estimate of $\beta(t)$ given by
\begin{align}
\hat \beta(t) = \sum_{j=1}^p \sum_{i=1}^N \frac{ \hat \xi_{ij} Y_i }{\hat \lambda_j N} \hat v_j(t), \label{e:pace:est}
\end{align}
where a hat denotes an arbitrary (at this point) estimate of the analogous quantity.  
The motivation for this estimator is the fact that $\E[ \xi_{ij} Y_i] = \lambda_j \langle \beta, v_j \rangle$.  However, as written, even if we knew the correct values for the parameters used for imputation, using PACE the most we could hope for would be
\[
\frac{1}{N}\sum_{i=1}^N { \hat \xi_{ij} Y_i }  \convP \E[ Y_i \E[ \xi_{ij} | \{x_{ik}\}]] \neq \E[ Y_i \xi_{ij}].
\]
That the last two terms are not equal implies that this estimate is biased unless either (1) $\beta(t) \equiv 0$ or  (2) the average number of points per curve tends to $\infty$ as the sample size increases.  In the former case, the equality holds since we would have $\E[Y_i \xi_{ij}] =  \E[Y_i] \E[\xi_{ij}]$ and $\E[ Y_i \E[ \xi_{ij} | \{x_{ik}\}]] = \E[Y_i] \E[\E[\xi_{ij} | \{x_{ik}\}]]= \E[Y_i] \E[\xi_{ij}]$, while in the latter case, one would have $\E[ \xi_{ij} | \{x_{ik}\}] \overset{P}{\to} \xi_{ij}$ as more points are collected (this of course requires additional assumptions).  Thus for very sparse, irregular designs the bias from plugging PACE BLUPs directly into subsequent estimation procedures can be meaningful.  

To alleviate this problem \cite{yao2005functionalR} avoided plugging the PACE BLUPs directly into \eqref{e:pace:est} and instead estimated $\E[Y_i \xi_{ij}]$ directly by smoothing the cross-covariance between $X_i(t)$ and $Y_i$.  While this approach works well in the linear case, it is very hard to extend to other settings when the parameter cannot be explicitly written down in terms of moments to be estimated.  
%

Our approach to fixing these issues is to utilize draws from the conditional distribution of $X_i(t)$ given both $Y_i$ and  $\{x_{ik}\}$, which is a form of multiple imputation.  In contrast, PACE is a form of mean imputation and does not condition on the outcome.  To carry out our multiple imputation framework, we need an imputation model that is \textit{compatible} with \eqref{e:sonf}.  That is, we need to ensure that our assumed distribution for $X_i(t) | Y_i, \{x_{ik}\}$ leads to the correct distribution for $Y_i | X_i$.  However, since all terms are assumed to be jointly Gaussian, it immediately follows that $X_i(t) | Y_i, \{x_{ik}\}$ is still a Gaussian process.  To carry out the conditional draws we therefore need only determine its mean and covariance.  
Assuming that $E[Y_i] = \mu_Y$ and $Var[Y_i] = \sigma_Y^2$ for $i=1,\cdots,N$, classic results on multivariate normality imply that
\begin{align*}
\E[X_i(t) | Y_i, \{x_{ik}\}] 
& = \mu_X(t) + \ba_i(t)^\top \bB_i  \bd_i  \\
\Cov(X_i(t), X_i(s)| Y_i, \{x_{ik}\}) 
&= C_X(t,s) - \ba_i(t)^\top \bB_i \ba_i(s),
\end{align*}
where
\[
\bd_i = 
\left(\begin{matrix}
Y_i - \mu_Y\\
x_{i1} - \mu_X(t_{i1})\\ 
\vdots \\
x_{im_i} - \mu_X(t_{im_i}) 
\end{matrix}\right)
\qquad
\ba_i(t) = \left(\begin{matrix}
C_{XY}(t) \\
C_X(t,t_{i1}) \\
\vdots \\
C_X(t, t_{im_i})
\end{matrix} \right)
\]
and 
\[
\bB_i^{-1}
 = \left(\begin{matrix}
\sigma_Y^2 & C_{XY}(t_{i1}) & C_{XY}(t_{i2})  & \dots \\
C_{XY}(t_{i1}) & C_X(t_{i1}, t_{i1}) + \sigma^2_\delta & C_X(t_{i1}, t_{i2}) & \dots \\
C_{XY}(t_{i2}) & C_X(t_{i2}, t_{i1}) & \ddots  & \vdots\\
\vdots & \dots & \dots & C_X(t_{im_i}, t_{im_i}) + \sigma^2_\delta 
\end{matrix}\right).
\]
Using these expressions we can, after estimating the requisite parameters, draw $K$ times from the above conditional distribution to form $K$ imputations, $X_i^{(k)}(t)$, $k=1,\dots,K$.   For each of these imputed samples we can form the complete data estimates, $\hat \beta^{(k)}(t)$, using any of a number of estimation methods (FPCA, splines, RKHS, etc).  In the simulations we will focus on FPCA-based estimates so as to better compare against PACE.

In the case where FPCA is going to be used to estimate $\beta(t)$, it can be convenient to impute the scores directly.  In this case, we are interested in the conditional distribution $\xi_{ij} | Y_i, \{x_{ik}\}$.  Nearly the same expressions can be used, except that the form for $\ba_i(t)$ changes into
\[
\bA_i = 
\left( \begin{matrix}
\langle C_{XY}, v_1 \rangle & \langle C_{XY}, v_2 \rangle   & \dots & \langle C_{XY}, v_p \rangle \\
\lambda_1 v_1(t_{i1}) & \lambda_2 v_2(t_{i1}) & \dots & \lambda_p v_p(t_{i1}) \\
\vdots \\
\lambda_1 v_1(t_{im_i})  & \lambda_2 v_2(t_{im_i}) & \dots & \lambda_p v_p(t_{im_i}) \\
\end{matrix} \right),
\]
and we then get that
\begin{equation}\label{eq:scores}
	\E[ \bxi_i | \bd_i ] = \bA_i^\top \bB_i \bd_i 
	\qquad
	\Var(\bxi_i | \bd_i) = \diag\{\lambda_1, \dots, \lambda_p\} -  \bA_i^\top \bB_i \bA_i.
\end{equation}
In this case, after the imputations are made, one can move directly to using the scores to estimate $\beta(t)$.   

A major advantage of using a multiple imputation approach is that we can account for the uncertainty introduced in the imputation process.  This is accomplished by using \textit{Rubin's rules} \citep{rubin2004multiple}, namely, we compute the estimated within- and between-imputation covariance functions of the $\hat \beta^{(k)}(t)$,
\[
\widehat W(t,s) = \frac{1}{K}\sum_{k=1}^K \widehat \Cov(\hat \beta^k(t), \hat \beta^k(s))
\qquad \widehat B(t,s) = \frac{1}{K-1} \sum_{k=1}^K (\hat \beta^{(k)}(t) - \bar \beta(t))(\hat \beta^{(k)}(s) - \bar \beta(s))
\]
and then use the following as our final estimates and estimated covariance functions
\[
\hat \beta(t) = \bar \beta(t) = \sum_{k=1}^K \hat{\beta}^{(k)}(t)
\qquad \widehat C_{\hat \beta}(t,s) = \widehat W(t,s) + (1+1/K) \widehat B(t,s).
\]
Using these quantities one can carry out statistical inference for $\beta(t)$.

\subsection{Categorical Outcomes}\label{sec:logistic}
In this section we describe how to extend our procedure to categorical outcomes from an exponential family.  In the next section we consider the continuous case, but a key takeaway is that the imputation is much simpler if the GLM to be fit uses a canonical link function, in which case imputation using Gaussian processes is justified.  With categorical outcomes, the canonical link is the logistic function. Thus, we focus on the case of binary logistic regression, though the arguments proceed analogously for multiple categories.   

Recall that a logistic regression model for $Y_i | X_i$ implies that $Y_i \in \{0,1\}$ and
\begin{equation}\label{eq:log}
\logit(p_i) = \alpha + \int X_i(t) \beta(t) \ dt,
\end{equation}
where $p_i  = \E[Y_i | X_i] = P(Y_i = 1 | X_i)$.
In order to extend our methodology we have to determine a proper imputation model for the $X_i(t)$.  In other words, we have to select a model for $X_i(t) | Y_i$ that implies that $Y_i | X_i(t)$ satisfies a logistic regression model.  Interestingly, under certain conditions, one can still assume that $X_i(t) | Y_i$ is Gaussian.  Such results come about in multivariate linear discriminant analysis when comparing to logistic regression for multivariate data \citep{efron1975efficiency}.  We can extend those concepts to the case of functional data and we end up with the theorem below, which is interesting in its own right.
\begin{theorem}\label{t:log}
Assume $Y_i \in \{0,1\}$ are iid $Bern(p_0)$ and $X_i \in L^2[0,1]$ with $X_i | Y_i =y \sim \mcN(\mu + y \Delta, C_X)$, where $C_X$ has full rank and $\Delta \in L^2[0,1]$.  
\begin{enumerate}
\item If $\| C_X^{-1/2} \Delta \|^2_{L^2} = \infty$, then $Y_i | X_i$ is degenerate.
\item If $\| C_X^{-1/2} \Delta \|^2_{L^2} < \infty$, but $\| C_X^{-1} \Delta \|^2_{L^2} = \infty$, then $Y_i | X_i$ is not degenerate, but \eqref{eq:log} cannot hold for any $\beta \in L^2[0,1]$.
\item If $\| C_X^{-1} \Delta \|^2_{L^2}< \infty$ then \eqref{eq:log} holds with $\beta = C_X^{-1} \Delta \in L^2[0,1]$.
\end{enumerate}
\end{theorem}
Before interpreting the above theorem, we stress one point about $C_X$ and $C_X^{-1}$.  Recall that $C_X$ is always a linear, self-adjoint compact operator, and its inverse will exist as long as the null space of $C_X$ only contains the zero function (i.e. all eigenvalues are positive).  However, even when $C_X^{-1}$ does exist, it is neither compact nor even bounded.   Thus, an implicit part of assuming that  $\| C_X^{-1/2} \Delta\|^2_{L^2}< \infty$ is the assumption that this quantity exists and is well-defined.  In fact, this condition can be extended to the case where $C_X$ has a nontrivial null space by using the Moore-Penrose generalized inverse and assuming that $\Delta$ is orthogonal to any element of said null space, though for ease of exposition we do not pursue that here.

Another subtle point is the ``grey area'' where $\| C_X^{-1/2} \Delta \|^2_{L^2}< \infty$, but $\| C_X^{-1} \Delta \|^2_{L^2}= \infty$.  In this case, the logistic regression exists and is nondegenerate, but one has to replace $\langle x , \beta\rangle$ with $T_\beta(x)$ where $T_\beta$ is a linear functional that is not continuous.  When $x$ is a sample path from $X$, $T_\beta(x)$ will be finite almost surely, but it need not be finite for a general element of $L^2[0,1]$.  Put another way, $T_\beta$ will be an element of the algebraic dual of $L^2[0,1]$, but not the topological dual.  

Theorem \ref{t:log} is based on the orthogonality/equivalence of probability measures.  In particular, if the distributions for $X_i(t)|Y_i=0$ and $X_i(t)|Y_i=1$ are orthogonal, then it is possible to determine the value of $Y_i$ from $X_i(t)$ with probability 1, and thus no logistic model can exist as the probabilities would have to be 0 or 1.  The quantity $\| C_X^{-1/2} \Delta \|^2_{L^2}$ comes up in both \citet{delaigle2012achieving} and \citet{dai2017optimal} in terms of classification for FDA.  They show there that if this condition is not satisfied then perfect classification is possible.  What was not discussed, however, was the connection to the orthogonality of Gaussian measures.  Indeed, this same quantity was discovered at least as early as the 70s (see the historical discussions in \cite{kuo1975gaussian,bogachev1998gaussian} for additional details).  Clearly, if two measures are orthogonal then it is possible to determine, with probability 1, whether a sample came from one or the other.  This issue was discussed more deeply in recent work by \citet{berrendero2018use} in the context of using RKHS methods for classification as well as \citet{mirshani2017existence} in the context of functional data privacy.

With these tools in hand, we can now carry out our imputation for logistic regression.  In particular, we simply impute the group for $Y_i=0$ and $Y_i=1$ separately (though common parameters are still estimated jointly, as discussed in Section \ref{sec:sims}):
\begin{align}\label{eq:logimp}
\E[X_i(t) | Y_i = y, \bx_i]
& = \mu_y(t) + \ba_i(t)^\top \bB_i (\bx_i - \bmu_i)\\
\Cov(X_i(t), X_i(s) | Y_i, \bx_i)
&=  C_X(t,s) - \ba_i(t)^\top \bB_i \ba_i(s), \notag
\end{align}
where we now have
\[
\ba_i(t) = \left(\begin{matrix}
C_X(t,t_{i1}) \\
\vdots \\
C_X(t, t_{im_i})
\end{matrix} \right),
\qquad
\bB_i^{-1}
= \left(\begin{matrix}
C_X(t_{i1}, t_{i1}) + \sigma^2_\delta & C_X(t_{i1}, t_{i2}) & \dots \\
C_X(t_{i2}, t_{i1}) & \ddots  & \vdots\\
\vdots & \dots & C_X(t_{im_i}, t_{im_i}) + \sigma^2_\delta 
\end{matrix}\right),
\]
$\bmu_i = \E[\bx_i | Y_i =y] =  \{\mu_y(t_{ij})\}$, and $\mu_y(t) $ is the mean of the $X_i$ from group $y$.  
In addition, one can do this for the scores as well.  However, a caveat is that one cannot use the scores directly in fitting the subsequent logistic model as the difference between the two groups is entirely captured by the means, which are removed when computing scores.  We thus recommend instead working with $\tilde \xi_{ij} = \langle \mu_y, v_j \rangle + \xi_{ij}$, which does not have the mean effect removed and still allows one to estimate the coefficients of $\beta(t)$ in the FPCA basis.  To impute the $\xi_{ij}$, much the same formula can be used but now
\[
\bA_i = 
\left( \begin{matrix}
\lambda_1 v_1(t_{i1}) & \lambda_2 v_2(t_{i1}) & \dots & \lambda_p v_p(t_{i1}) \\
\vdots \\
\lambda_1 v_1(t_{im_i})  & \lambda_2 v_2(t_{im_i}) & \dots & \lambda_p v_p(t_{im_i}) \\
\end{matrix} \right),
\]
is used in place of $\ba_i(t)$ in \eqref{eq:logimp}, and $\mu_y(t)$ and $C_X(t,s)$ are replaced by the mean and covariance of the scores---\textbf{0} and $\text{diag}(\lambda_1,\cdots,\lambda_p)$, respectively.  Again, after imputation one should then construct $\tilde \xi_{ij}$ which have the means added back in, before fitting the subsequent logistic model.

\subsection{Continuous Outcomes} \label{sec:cont}
Lastly we consider the case where $Y$ is continuous, but not necessarily Gaussian.  As with the previous section we will show that a Gaussian imputation model for $X | Y$ will be consistent with a GLM for $Y|X$, however we require that $Y|X$ utilize a canonical link function.  In particular, this means that the conditional density of $Y|X=x$ is given by
\begin{align}
f_{Y|X}(y| x) =  h(y) \exp\{ \eta_x T(y) - A(\eta_x) \}, \label{eq:glm}
\end{align}
where $\eta_x = \alpha + \langle \beta, x\rangle$, $h$ is some known function that determines the family of the GLM, $T$ is a known function representing the sufficient statistic, and $\exp\{-A(\eta_x)\}$ is the normalizing constant.

\begin{theorem}\label{t:glm}
Assume $Y_i \in \mbR$ are iid with density $f(y) \propto h(y) \exp\{- (1/2) T(y)^2 \| C_X^{-1/2}\Delta\|^2 \}$, which is assumed to be integrable, $X_i \in L^2[0,1]$ with $X_i | Y_i =y \sim \mcN(\mu + T(y) \Delta, C_X)$, and $\Delta \in L^2[0,1]$.  
\begin{enumerate}
\item If $\| C_X^{-1/2} \Delta \|^2_{L^2} = \infty$, then $Y_i | X_i$ is degenerate.
\item If $\| C_X^{-1/2} \Delta \|^2_{L^2} < \infty$, but $\| C_X^{-1} \Delta \|^2_{L^2} = \infty$, then $Y_i | X_i$ is not degenerate, but \eqref{eq:glm} cannot hold for any $\beta \in L^2[0,1]$.
\item If $\| C_X^{-1} \Delta \|^2_{L^2}< \infty$ then \eqref{eq:glm} holds with $\beta = C_X^{-1} \Delta \in L^2[0,1]$. 
\end{enumerate}
\end{theorem}

In extending to more general exponential families, the key hurdle appears to be working with a noncanonical link function.  When swapping from conditioning $Y|X$ to $X|Y$, the latter is always in a canonical form.  While on the surface this may seem minor, this appears to be a serious open mathematical question.  It is desirable to assume that $X|Y$ is Gaussian since this simplifies the imputation, however, then one is essentially ``boxed in'' to using a canonical link function.

\subsection{Computation}\label{sub:comp}

Implementing MISFIT for either linear or logistic regression requires the estimation of a number of parameters in the imputation model. While the estimation of these imputation parameters is not our focus, we wish to make clear how one can implement our imputation strategy. To do so, we dedicate this section to outlining computation of estimates for the imputation parameters, connecting back to the results from sections \ref{sub:lm} and \ref{sec:logistic}.

The results provided in Sections \ref{sec:sims} and \ref{sec:app} were computed in {\tt R} \citep{R}, using the package {\tt fcr} \citep{fcr} to estimate imputation parameters. Based on \citet{leroux2017dynamic}, this package is designed to fit functional concurrent regression models and allows us to regress $X_i(t)$ on $Y_i$ as follows:
\begin{equation}\label{eq:fcr}
X_i(t) = f_0(t) + f_1(t)Y_i + b_i(t),
\end{equation}
where the $b_i(t)\overset{iid}{\sim} N(0,C_b(t,s))$ are curve-specific random effects. The curves are observed with noise such that
\begin{equation}\label{eq:fcrobs}
X_i(t_{ij}) = f_0(t_{ij}) + f_1(t_{ij})Y_i + b_i(t_{ij}) + \delta_{ij},
\end{equation}
with the $b_i(t_{ij})$ and $\delta_{ij}$ mutually independent. Through fitting this model, we obtain estimates of $f_0(t)$, $f_1(t)$, $C_b(t,s)$, and $\sigma_{\delta}^2$, from which we can in turn estimate all necessary imputation parameters.\\

\noindent\textbf{Linear Model:}
Using equations \eqref{eq:fcr} and \eqref{eq:fcrobs}, we can directly compute the mean and covariance of $X_i(t)$ as well as the cross covariance between $X_i(t)$ and $Y_i$. This gives us:
\begin{align*}
\mu_X(t) & = f_0(t) + f_1(t)\mu_Y   &   
\mu_X(t_{ij}) & = f_0(t_{ij}) + f_1(t_{ij})\mu_Y\\
C_X(t,s) & = f_1(t)f_1(s)\sigma_Y^2 + C_b(t,s)   &   
C_X(t_{ij},s) & = f_1(t_{ij})f_1(s)\sigma_Y^2 + C_b(t_{ij},s)\\
C_{XY}(t) & = f_1(t)\sigma_Y^2   &   C_{XY}(t_{ij}) & = f_1(t_{ij})\sigma_Y^2
\end{align*}
and 
\[
C_X(t_{ij},s_{ik}) = f_1(t_{ij})f_1(s_{ik})\sigma_Y^2 + C_b(t_{ij},s_{ik}) + 1_{\{j=k\}}\sigma_{\delta}^2.
\]
From there, one can obtain the $\lambda_j$ and $v_j(t)$ from a spectral decomposition of $C_X(s,t)$.\\

\noindent\textbf{Logistic Model:}
As our logistic regression imputation described in \eqref{eq:logimp} is done separately for the two groups (i.e. conditioning on $Y_i$), the results are somewhat different than in the linear case and contain no parameters of $Y_i$. That is, we instead compute $E[X_i(t)|Y_i=y] = \mu_y(t)$ and $Cov(X_i(t),X_i(s)|Y_i=y) = C_X(t,s)$. Again, from equations \eqref{eq:fcr} and \eqref{eq:fcrobs} we get
\begin{align*}
\mu_y(t) & = f_0(t) + f_1(t)Y_i   &   \mu_y(t_{ij}) & = f_0(t_{ij}) +f_1(t_{ij})Y_i\\
C_X(t,s) & = C_b(t,s)   &   C_X(t_{ij},s) & = C_b(t_{ij},s)
\end{align*} 
and
\[
C_X(t_{ij},s_{ik}) = C_b(t_{ij},s_{ik}) + 1_{\{j=k\}}\sigma_{\delta}^2.
\]
Again, we can readily compute the $\lambda_j$ and $v_j$ from $C_X(t,s)$, and that gives us all that we need since the logistic imputation model does not require $C_{XY}$, $\mu_Y$, or $\sigma_Y^2$.


\section{Simulations}\label{sec:sims}
So far, we have advocated for imputing the curves (scores) of sparse functional data by forming and drawing multiple times from the conditional distribution of the curves (scores) given the observed values of the response variable and the observed points of the predictor curve. We have posed this method, MISFIT, as an alternative to the PACE method, which differs in that it does not condition on the response variable, and it imputes solely based on the mean of the conditional distribution. For the sake of brevity, we will refer to these methods as the Multiple Conditional (MuC) and the Mean Unconditional (MeU) approach, respectively. In addition to comparing these two methods in simulations, we find it enlightening to also compare them against their intermediary counterparts--the Mean Conditional (MeC) and Multiple Unconditional (MuU) approaches. 

We compare these four approaches in both a linear and logistic scalar-on-function regression setting, investigating the estimation accuracy, as well as the type 1 error rates and power, of their resulting estimators. Since we expect that the MeU imputation approach is biased for a small average number of observations per curve, $m$, we compare across simulated data sets with varying values of $m$, as well as varying sample sizes. In addition, since the FPCA-estimator given in \ref{e:pace:est} depends on the value $J$ to truncate the sum, we must specify a fixed $J$ for all imputation approaches with each simulated data set. For the multiple imputation approaches (MuU and MuC) we generated $K = 10$ completed datasets for all of the following simulations. Finally, while in section \ref{sec:meth} we detailed both a curve-level and a score-level imputation strategy, we use only the score-level imputation in all of the simulations.

\subsection{Linear Model}\label{linear_sim}

For a linear model, we first simulate $N$ iid random curves, $\{X_1(t),\cdots,X_N(t)\}$, from a mean-0 Gaussian process with the following Mat\'{e}rn covariance function

\begin{equation}\label{eq:matern}
C_X(t,s) = \frac{\sigma^2}{\Gamma(\nu)2^{\nu - 1}}\left(\frac{\sqrt{2\nu}|t-s|}{\rho}\right)^\nu K_{\nu}\left(\frac{\sqrt{2\nu}|t-s|}{\rho}\right),
\end{equation}

\noindent where $K_{\nu}$ represents the modified Bessel function of the second kind, and we choose $\nu = 5/2$, $\rho = 0.5$, and $\sigma^2 = 1$. These curves are evaluated at $M=100$ equally-spaced times from $[0,1]$. Since we assume that the random curves are observed with error, we add noise to the realized curves to produce the observed curves, where $\sigma_\delta^2 = 0.5$. We next define $\beta(t) = w\times\text{sin}(2\pi t)$, where $w$ is a weight coefficient chosen to adjust the signal. The responses, $Y_i$, $i=1,\cdots,N$, are then generated according to model \ref{e:sonf}, where $\alpha = 0$ and $\sigma_\epsilon^2 = 1$. Finally, for each observed curve, we randomly sample $m$ time points from the length-100 grid to observe, so that the observed data used for imputation is $\{Y_i,x_{i1},\cdots,x_{im} \},$ for $i = 1,\cdots,N$. Once the scores are imputed, a linear regression model is fit using the first $J$ scores.

\subsubsection{True Parameters}\label{true_sim}

For the first set of simulations, we treat the true parameters as known; that is, we use the true values for $\sigma_Y^2$, $C_{XY}(t)$, $C_X(s,t)$, $\sigma_\delta^2$, $v_j(t)$, and $\lambda_j$, for $j=1,\cdots,J$. In addition, we simulate data sets of different sample sizes, $N \in \{100, 200, 400, 800\}$; different number of observations per curve, $m \in \{2, 5, 10, 20\}$; different numbers of FPCs, $J \in \{1,\cdots,6\}$; and with different signals, $w \in \{0, 5, 10\}$. Each of these settings is simulated 1,000 times. Since we are primarily interested in accuracy of the final estimates $\hat{\beta}(t)$, we report the  median integrated squared error (MISE) of $\hat{\beta}(t)$, defined as $median \int(\hat{\beta}_s(t) - \beta(t))^2dt$, where $s=1,\cdots,1000$ indexes the particular run and the median is taken over these 1,000 simulations. The \textit{median} ISE is reported instead of the \textit{mean} ISE because, particularly when $m = 2$ or $5$, the imputation methods are prone to occasional large outliers. 

Tables \ref{tab:trueJ}, \ref{tab:trueN}, and \ref{tab:truem} show the MISE of $\hat{\beta}(t)$ 
across varying $N$, $m$, and $J$, respectively. For all of the simulations, the default settings of $N = 200$, $m = 2$, and $J = 4$ are used, allowing one of these to vary in each table (e.g. in table \ref{tab:trueN} $m = 2$ and $J = 4$, while $N$ varies). The choice of $J = 4$ will become obvious after consideration of table \ref{tab:trueJ}. However we also ran simulations with $J = 3$ and $J = 5$, the results of which can be found in the appendix, section \ref{sec:apptables}. The comparisons are much the same, regardless of the choice of $J$.

Beginning with a look at table \ref{tab:trueJ}, notice that when the sample size is small ($N = 200$) and curves are observed very sparsely ($m = 2$), regardless of how many FPCs are used and the strength of the signal, MuC is the most accurate method. In the presence of some signal, the 
MISE for MuC can be reasonably large when too few (e.g. 1, 2, or 3) FPCs are used, driven by truncation bias. In these cases, MuC's advantage over the other methods is most accentuated when $J$ is chosen as 4 or 5, in which case enough FPCs are used to capture the complexity of the shape of $\beta(t)$. It is worth highlighting that, in the absence of parameter estimation error, choosing a large enough value $J$ results in an (approximately) unbiased estimator for MuC. However, the variance of the estimator for MuC tends to increase with larger choices of $J$, so some balance is required. 
Based on our observations, the variance of the MuC estimator remains relatively low for all choices of $J$, while the variance of the mean-imputation-based estimators balloons for larger $J$. The MuU estimator typically exhibited less variance than the mean-imputation-based estimators did, but tended to oversmooth its estimate, yielding quite a large bias in the presence of stronger signals.

\begin{table}[ht]
\centering
\scalebox{0.9}{
\begin{tabular}{lllllllllllll}
  \hline
   J & \multicolumn{4}{c}{w = 0} & \multicolumn{4}{c}{w = 5} & \multicolumn{4}{c}{w = 10} \\ \cmidrule(lr){2-5}\cmidrule(lr){6-9}\cmidrule(lr){10-13}
  & MeC & MuC & MeU & MuU & MeC & MuC & MeU & MuU & MeC & MuC & MeU & MuU \\
 \hline
1 & 0.00 & 0.00 & 0.00 & 0.00 & 12.39 & 12.38 & 12.39 & 12.38 & 49.54 & 49.52 & 49.55 & 49.52 \\ 
  2 & 0.05 & 0.01 & 0.05 & 0.01 & 4.95 & 3.30 & 3.40 & 7.68 & 13.83 & 13.16 & 13.46 & 30.80 \\ 
  3 & 0.74 & 0.02 & 0.74 & 0.02 & 5.26 & 3.31 & 5.02 & 7.73 & 13.96 & 13.17 & 17.99 & 30.95 \\ 
  4 & 12.09 & 0.06 & 12.09 & 0.06 & 138.17 & 0.10 & 28.24 & 7.55 & 43.55 & 0.23 & 77.33 & 30.14 \\ 
  5 & 177.30 & 0.19 & 177.30 & 0.19 & 198.40 & 0.21 & 430.48 & 8.07 & 59.15 & 0.34 & 1125.15 & 31.71 \\ 
  6 & 2149.97 & 0.57 & 2149.97 & 0.57 & 764.15 & 0.53 & 5096.74 & 9.35 & 205.31 & 0.58 & 13141.37 & 35.26 \\ 
   \hline
\end{tabular}
}
\caption{MISE of $\hat{\beta}(t)$ as $J$ increases,
                          for a linear model using true imputation parameters.} 
\label{tab:trueJ}
\end{table}
In table \ref{tab:trueN} we see that increasing the sample size benefits all four approaches. The improvement in MISE occurs through a reduction in variance rather than a reduction in bias. The MuC estimator noticeably has the smallest MISE across all signals and all sample sizes. Notice that the tendency of MuC's MISE towards zero gives evidence that MuC results in a consistent estimator. Its minimal MISE at larger sample sizes is due to truncation error resulting from the use of a finite number of FPCs; as table \ref{tab:trueJ} showed, a larger choice of $J$ helps to alleviate this for MuC. 

The second best estimator is typically MuU, except in the absence of a signal, in which case MuU and MuC are equivalent. On the other hand, the large and persistent 
MISE of the MeC and MuU estimators as the sample size grows is noteworthy. 
MeU clearly does poorly for small sample sizes, especially when the signal is strong. And while consistency for MeU has already been established by \citet{yao2005functionalR}, MuC clearly outperforms MeU in such a sparse setting, even with a large sample size.

\begin{table}[!h]
\centering
\scalebox{1}{
\begin{tabular}{lllllllllllll}
  \hline
   N & \multicolumn{4}{c}{w = 0} & \multicolumn{4}{c}{w = 5} & \multicolumn{4}{c}{w = 10} \\ \cmidrule(lr){2-5}\cmidrule(lr){6-9}\cmidrule(lr){10-13}
  & MeC & MuC & MeU & MuU & MeC & MuC & MeU & MuU & MeC & MuC & MeU & MuU \\
 \hline
100 & 28.24 & 0.13 & 28.24 & 0.13 & 139.05 & 0.17 & 64.86 & 7.71 & 43.35 & 0.28 & 175.47 & 30.72 \\ 
  200 & 12.09 & 0.06 & 12.09 & 0.06 & 138.17 & 0.10 & 28.25 & 7.55 & 43.54 & 0.23 & 77.37 & 30.14 \\ 
  400 & 5.45 & 0.03 & 5.45 & 0.03 & 139.33 & 0.07 & 14.16 & 7.58 & 43.43 & 0.21 & 40.06 & 30.30 \\ 
  800 & 2.89 & 0.02 & 2.89 & 0.02 & 138.29 & 0.06 & 6.54 & 7.49 & 43.36 & 0.19 & 19.22 & 30.02 \\ 
   \hline
\end{tabular}
}
\caption{MISE of $\hat{\beta}(t)$ as $N$ increases,
                          for a linear model using true imputation parameters.} 
\label{tab:trueN}
\end{table}
The previous two tables have showed convincing results that MuC is the best estimator when $m = 2$, as we expected. Table \ref{tab:truem} allows us to compare across the four approaches for an increasing number of observed points per curve, $m$. Unsurprisingly, the four approaches begin to converge in MISE, regardless of the strength of the signal, as $m$ increases and the amount of information observed prior to imputation grows. Still, MuC maintains its status of lowest MISE even up to 20 observations per curve. 

One other interesting point borne out by table \ref{tab:truem} is that MuC results in a lower MISE the sparser the observed curves. Since MuC draws imputed values from the correct theoretical distribution, multiple imputation results in multiple sets of ``correct" scores. 
Thus, imputing more scores actually decreases the variability of the resulting estimate; the increasing MISE of MuC as $m$ grows is purely an artifact of this phenomenon. We will see in the next section that using estimated imputation parameters as opposed to the true parameters disturbs the distribution used by MuC enough to nullify this aberration.

\begin{table}[!h]
\centering
\scalebox{1}{
\begin{tabular}{lllllllllllll}
  \hline
   m & \multicolumn{4}{c}{w = 0} & \multicolumn{4}{c}{w = 5} & \multicolumn{4}{c}{w = 10} \\ \cmidrule(lr){2-5}\cmidrule(lr){6-9}\cmidrule(lr){10-13}
  & MeC & MuC & MeU & MuU & MeC & MuC & MeU & MuU & MeC & MuC & MeU & MuU \\
 \hline
2 & 12.09 & 0.06 & 12.09 & 0.06 & 138.19 & 0.10 & 28.25 & 7.55 & 43.55 & 0.23 & 77.37 & 30.14 \\ 
  5 & 3.89 & 0.10 & 3.89 & 0.10 & 67.63 & 0.13 & 7.65 & 4.12 & 32.47 & 0.24 & 17.73 & 16.45 \\ 
  10 & 1.82 & 0.14 & 1.82 & 0.14 & 30.47 & 0.16 & 2.64 & 2.69 & 23.54 & 0.27 & 5.71 & 10.31 \\ 
  20 & 1.08 & 0.19 & 1.08 & 0.19 & 10.84 & 0.22 & 1.45 & 1.83 & 14.09 & 0.31 & 2.41 & 6.95 \\ 
   \hline
\end{tabular}
}
\caption{MISE of $\hat{\beta}(t)$ as $m$ increases, 
                          for a linear model using true imputation parameters.} 
\label{tab:truem}
\end{table}

\subsubsection{Estimated Parameters}\label{est_sim}

Unlike in the previous section, in practice the imputation parameters must themselves first be estimated. Here, we mimic the above simulations replacing the true imputation parameters, $\sigma_Y^2$, $C_{XY}(t)$, $C_X(s,t)$, $\sigma_\delta^2$, $v_j(t)$, and $\lambda_j$, for $j=1,\cdots,J$, with their estimates from the data.

For the two imputation approaches that condition on the outcome, we obtain estimates $\hat{C}_{XY}(t)$, $\hat{C}_X(s,t)$, $\hat{\mu}(t)$, and $\hat{\sigma}_\delta^2$ using {\tt fcr} \citep{fcr} to regress $X_i(t)$ on $Y_i$. Estimates $\{\hat{\lambda}_j\}_{j=1}^J$ and $\{\hat{v}_j(t)\}_{j=1}^J$ are then taken to be the eigenvalues and eigenfunctions of $\hat{C}_X(t)$, and the usual sample mean and variance are used for $\hat{\mu}_Y$ and $\hat{\sigma}_Y^2$. For the unconditional imputation approaches, we ignore the $Y_i$ and use the function {\tt face.sparse} \citep{xiao2018fast} from the {\tt face} package \citep{face} to compute estimates $\hat{C}_X$, $\hat{\mu}(t)$, and $\hat{\sigma}_\delta$ (where again $\{\hat{\lambda}_j\}_{j=1}^J$ and $\{\hat{v}_j(t)\}_{j=1}^J$ are obtained from a spectral decomposition of $\hat{C}_X$).

As in the previous simulations, two of $N$, $m$, and $J$ are fixed while 
the third varies. When a value isn't varying we take defaults of 
$N=200$, $m=2$, and $J=2$. The different settings considered are $N \in \{ 100, 200, 400, 800 \}$, $m \in \{ 2, 5, 10, 20\}$, and $J \in \{ 1,\cdots,6\}$, and simulations under each setting are performed 100 times. Tables \ref{tab:estJ}, \ref{tab:estN}, and \ref{tab:estm}, respectively, are the analogues to tables \ref{tab:trueJ}, \ref{tab:trueN}, and \ref{tab:truem} above, showing the MISE of $\hat{\beta}(t)$ as $N$, $m$, and $J$ increase. In table \ref{tab:estJ} we see that, except when $\beta(t)\equiv 0$ (i.e. $w = 0$), in which case choosing the smallest number of FPCs clearly makes the most sense, a choice of $J = 2$ is otherwise the most reasonable. 

Table \ref{tab:estJ} is noticeably different in contrast to table \ref{tab:trueJ}, where we used the true imputation parameters. 
Using estimated imputation parameters instead, MuC has noticeable MISE across almost all values of $J$, and in particular does not become unbiased by simply increasing $J$. This results from bias in estimating the imputation parameters when both $m$ and $N$ are small. In these simulations, MuC no longer boasts a distinctive advantage in MISE across all values of $J$, although it still outperforms the other methods when the signal is large. The best choice of $J$, therefore, again depends on the signal and is not unanimous, but a choice of $J = 2$ seems the most equitable for the four approaches.

\begin{table}[h]
\centering
\scalebox{0.85}{
\begin{tabular}{lllllllllllll}
  \hline
   J & \multicolumn{4}{c}{w = 0} & \multicolumn{4}{c}{w = 5} & \multicolumn{4}{c}{w = 10} \\ \cmidrule(lr){2-5}\cmidrule(lr){6-9}\cmidrule(lr){10-13}
  & MeC & MuC & MeU & MuU & MeC & MuC & MeU & MuU & MeC & MuC & MeU & MuU \\
 \hline
1 & 0.02 & 0.01 & 0.01 & 0.00 & 12.36 & 12.35 & 12.35 & 12.36 & 49.40 & 49.36 & 49.38 & 49.42 \\ 
  2 & 1.98 & 0.11 & 0.11 & 0.01 & 7.61 & 6.15 & 9.24 & 8.36 & 21.22 & 20.96 & 39.37 & 33.13 \\ 
  3 & 656.22 & 1.74 & 71.87 & 0.09 & 30.39 & 12.89 & 260.60 & 9.06 & 30.41 & 28.38 & 855.85 & 35.11 \\ 
  4 & 9103.84 & 64.31 & 5298.38 & 0.86 & 50.98 & 36.56 & 20714.05 & 11.92 & 67.35 & 55.86 & 41600.57 & 43.91 \\ 
  5 & 43224.30 & 264.08 & 360950.11 & 5.38 & 96.31 & 71.24 & 926620.26 & 26.28 & 100.43 & 84.95 & 2182182.21 & 91.21 \\ 
   \hline
\end{tabular}
}
\caption{MISE of $\hat{\beta}(t)$ as $J$ increases,
                          for a linear model using estimated imputation parameters.} 
\label{tab:estJ}
\end{table}
We can see in table \ref{tab:estN} that when $w = 0$, the unconditional approaches are more accurate than the conditional approaches, but that the advantage fades as the sample size grows. Conversely, as the signal increases, the advantage flips to the conditional approaches, particularly for small sample sizes. In particular, notice that with a sample size of 100, the MISE for MeU is about double that of MuC when $w = 5$ or $w = 10$. While these gains diminish as $N$ increases, as long as the signal is large enough, MuC results in the lowest MISE for all values of $N$, and its advantage increases with the signal. 

\begin{table}[h]
\centering
\scalebox{1}{
\begin{tabular}{lllllllllllll}
  \hline
   N & \multicolumn{4}{c}{w = 0} & \multicolumn{4}{c}{w = 5} & \multicolumn{4}{c}{w = 10} \\ \cmidrule(lr){2-5}\cmidrule(lr){6-9}\cmidrule(lr){10-13}
  & MeC & MuC & MeU & MuU & MeC & MuC & MeU & MuU & MeC & MuC & MeU & MuU \\
 \hline
100 & 2.77 & 0.16 & 0.24 & 0.02 & 7.40 & 6.90 & 11.95 & 8.85 & 21.64 & 22.37 & 47.26 & 35.08 \\ 
  200 & 1.90 & 0.11 & 0.11 & 0.01 & 7.61 & 6.15 & 9.54 & 8.31 & 21.22 & 20.94 & 39.59 & 33.03 \\ 
  400 & 0.56 & 0.04 & 0.03 & 0.00 & 7.13 & 5.37 & 7.02 & 8.09 & 17.47 & 16.92 & 28.04 & 32.51 \\ 
  800 & 0.29 & 0.02 & 0.02 & 0.00 & 6.78 & 4.74 & 5.68 & 8.09 & 15.56 & 15.17 & 22.55 & 32.27 \\ 
   \hline
\end{tabular}
}
\caption{MISE of $\hat{\beta}(t)$ as $N$ increases,
                          for a linear model using estimated imputation parameters.} 
\label{tab:estN}
\end{table}
Table \ref{tab:estm} shows a similar pattern, where the conditional approaches overtake the unconditional approaches as the signal grows, and especially for smaller values of $m$. Again, for a large enough signal, regardless of the value of $m$, MuC outperforms each of the other imputation approaches. In addition, recall that in table \ref{tab:truem}, in which the true imputation parameters were used for imputation, the 
MISE of the MuC estimator actually increased slightly as more of each curve was observed. This is no longer the case since now increasing $m$ improves the imputation parameter estimates, leading to a better approximation of the imputation distribution.

\begin{table}[h]
\centering
\scalebox{1}{
\begin{tabular}{lllllllllllll}
  \hline
   m & \multicolumn{4}{c}{w = 0} & \multicolumn{4}{c}{w = 5} & \multicolumn{4}{c}{w = 10} \\ \cmidrule(lr){2-5}\cmidrule(lr){6-9}\cmidrule(lr){10-13}
  & MeC & MuC & MeU & MuU & MeC & MuC & MeU & MuU & MeC & MuC & MeU & MuU \\
 \hline
2 & 1.90 & 0.11 & 0.11 & 0.01 & 7.61 & 6.08 & 9.10 & 8.36 & 21.24 & 20.96 & 37.76 & 33.13 \\ 
  5 & 0.08 & 0.03 & 0.02 & 0.01 & 5.80 & 4.57 & 5.04 & 5.54 & 16.06 & 15.50 & 20.00 & 22.19 \\ 
  10 & 0.03 & 0.02 & 0.02 & 0.01 & 4.62 & 3.97 & 4.62 & 4.77 & 14.92 & 14.33 & 18.43 & 19.07 \\ 
  20 & 0.02 & 0.02 & 0.02 & 0.02 & 3.88 & 3.73 & 4.05 & 4.11 & 14.21 & 14.08 & 16.23 & 16.39 \\ 
   \hline
\end{tabular}
}
\caption{MISE of $\hat{\beta}(t)$ as $m$ increases,
                          for a linear model using estimated imputation parameters.} 
\label{tab:estm}
\end{table}

\subsubsection{Comparison of MISFIT, PACE, pfr, and funreg}\label{pfr_comparison}

In addition to comparing MISFIT's performance against the above intermediary imputation-based approaches, an obvious benchmark is the performance of other methods used for performing functional regression. While the above simulations also compare MISFIT to PACE, the {\tt fdapace} package \citep{fdapace} contains an implementation of PACE which uses local polynomial smoothing to estimate imputation parameters. Other useful benchmarks include {\tt pfr} of the package {\tt refund} \citep{refund}, a popular and flexible tool for fitting a wide range of functional regression models, and {\tt funreg} \citep{funreg}, which is specifically designed to accommodate functional regression with irregularly-spaced designs. The same simulation set-up as described above in section \ref{est_sim} was followed and 100 simulations were performed under each setting, using different values of $m$ only.

Table \ref{tab:compare} shows the MISE of MISFIT, PACE, pfr, and funreg across the different simulation settings. Of primary importance is the comparison of PACE and MISFIT. It is again evident that MISFIT outperforms PACE in nearly every setting and remains competitive where it does not. Also worth highlighting is that the results displayed for PACE in table \ref{tab:compare} are slightly different than those displayed in \ref{tab:estm} for MeU; these differences are purely due to differences in estimating the imputation parameters. The performances of {\tt pfr} and especially {\tt funreg} are noticeably worse than both PACE and MISFIT when $m$ is as small as 2 or 5. This is unsurprising as neither are by default designed to handle sparse functional data; in fact, {\tt funreg} is designed specifically for intensive longitudinal data. Also unsurprisingly, then, these two outperform PACE and MISFIT when the signal is strong and each function is observed at least as many as 20 times.

\begin{table}[h]
\centering
\scalebox{0.85}{
\begin{tabular}{lllllllllllll}
  \hline
   m & \multicolumn{4}{c}{w = 0} & \multicolumn{4}{c}{w = 5} & \multicolumn{4}{c}{w = 10} \\ \cmidrule(lr){2-5}\cmidrule(lr){6-9}\cmidrule(lr){10-13}
  & MISFIT & PACE & Pfr & Funreg & MISFIT & PACE & Pfr & Funreg & MISFIT & PACE & Pfr & Funreg \\
 \hline
2 & 0.11 & 0.04 & 0.57 & 172668.29 & 6.08 & 6.98 & 10.64 & 420328.17 & 20.96 & 27.98 & 40.62 & 1315775.00 \\ 
  5 & 0.03 & 0.02 & 0.05 & 1.58 & 4.57 & 4.19 & 5.27 & 10.13 & 15.50 & 16.75 & 21.00 & 34.99 \\ 
  10 & 0.02 & 0.02 & 0.04 & 0.62 & 3.97 & 4.02 & 5.11 & 5.67 & 14.33 & 16.00 & 19.56 & 21.07 \\ 
  20 & 0.02 & 0.02 & 0.04 & 0.23 & 3.73 & 3.74 & 5.07 & 5.12 & 14.08 & 14.88 & 13.42 & 8.72 \\ 
   \hline
\end{tabular}
}
\caption{Comparison of MISE when estimating $\beta(t)$ via MISFIT, Pace, pfr, and funreg, using estimated imputation parameters.} 
\label{tab:compare}
\end{table}

 \subsubsection{Type 1 Error Rates}\label{rej_rates}
As mentioned in section \ref{sub:lm}, one of the advantages of multiple imputation is that we can incorporate missing data uncertainty into estimation and statistical inference, which is neglected by mean imputation. As such, we would expect mean imputation to produce artificially small standard errors. We substantiate these expectations by simulating data, testing the hypothesis 
\[
H_0: \beta(t)\equiv 0
\] 
against the alternative 
\[
H_1: \beta(t)\not\equiv 0
\]
and comparing rejection rates across imputation methods. If, as we expect, mean imputation approaches underrepresent the standard error, then we should observe higher rejection rates for mean imputation approaches than their multiple imputation alternatives (i.e. a gain in power, but also larger type 1 error rates).

We again use estimated parameters and follow the same simulation and estimation procedures outlined above in section \ref{est_sim}. Hypotheses are tested at the 0.05 nominal significance level, using the statistic 
\[
T=||\beta||^2 \sim \sum_i^\infty \lambda_i^{*}\chi^2_i(1),
\]
where the $\lambda_i^{*}$ are the eigenvalues of $C_\beta(t,s)$. P-values are computed using the {\tt imhof()} function of the {\tt CompQuadForm} package. These simulations are run 500 times for each of three different signals, $w \in \{ 0, 1, 2\}$, and across all runs $N=200$ and $J=2$ are fixed, while $m \in \{ 2, 5, 10, 20 \}$. Note that simulations for $w = 0$ correspond to simulations under the null hypothesis and thus we would hope for our imputation method to have an empirical rejection rate close to the nominal Type 1 error rate of 0.05. Rejection rates for $w = 1$ and $w = 2$ correspond to statistical power. 

Table \ref{tab:reject} displays the average rejection rate for each imputation method. 
We immediately see that, as expected, rejection rates for the mean imputation approaches are much larger than those of their corresponding (i.e. conditional or unconditional) multiple imputation approaches until the signal becomes large enough and sufficiently many points per curve are observed. In the sparsest cases--when we only observe 2 or 5 points per curve--rejection rates for mean imputation can be substantially larger than those for multiple imputation. These differences tend to dissipate as $m$ increases to 20 points per curve, at which point all methods are fairly comparable.

It is interesting to compare the conditional imputation approaches to the unconditional ones. Clearly MuU and MeU perform better under the null hypothesis, while MuC and MeC are better at detecting true signals. The gain in power for MuC and MeC over MuU and MeU is quite dramatic when the signal is small ($w=1$) and the observations are especially sparse ($m=2$). It is also noteworthy that, while MeU appears to be the most calibrated to the nominal rate under the null hypothesis, the empirical rejection rate of MuC (and the other approaches) tends towards the nominal rate as the curves are observed more densely.

In summary, mean imputation leads to higher rejection rates than multiple imputation, as expected. However, one additional not-so-obvious result is that when the covariate is observed very sparsely, imputing the covariate conditional on the response inflates the type I error rate beyond the nominal rate. This makes sense, though, as the conditional imputation approach incorporates information from the response into the imputed values of the covariate; any false signal detected is merely residue of this process. The takeaway, then, is that one should only use the conditional imputation approach if there is evidence a priori (i.e. before imputation) that the variables are related, or for larger values of $m$. 
\begin{table}[h]
\centering
\begin{tabular}{lllllllllllll}
  \hline
   m & \multicolumn{4}{c}{w = 0} & \multicolumn{4}{c}{w = 5} & \multicolumn{4}{c}{w = 10} \\ \cmidrule(lr){2-5}\cmidrule(lr){6-9}\cmidrule(lr){10-13}
  & MeC & MuC & MeU & MuU & MeC & MuC & MeU & MuU & MeC & MuC & MeU & MuU \\
 \hline
2 & 0.704 & 0.318 & 0.050 & 0.000 & 0.928 & 0.786 & 0.470 & 0.000 & 0.960 & 0.968 & 0.890 & 0.116 \\ 
  5 & 0.302 & 0.118 & 0.062 & 0.000 & 0.962 & 0.892 & 0.834 & 0.398 & 1.000 & 1.000 & 1.000 & 0.972 \\ 
  10 & 0.144 & 0.056 & 0.042 & 0.012 & 0.974 & 0.928 & 0.920 & 0.792 & 1.000 & 1.000 & 1.000 & 1.000 \\ 
  20 & 0.082 & 0.056 & 0.050 & 0.036 & 0.968 & 0.938 & 0.938 & 0.904 & 1.000 & 1.000 & 1.000 & 1.000 \\ 
   \hline
\end{tabular}
\caption{Rejection rates at the 0.05 significance level over 500 simulations.} 
\label{tab:reject}
\end{table}

\subsection{Multivariate T Distribution}\label{tdist_sim}

One key assumption of MISFIT's imputation strategy for linear models is that the functional covariate, $X(t)$, follows a Gaussian process. To assess MISFIT's robustness to this assumption, we instead simulate the random sample of functions from a multivariate t distribution with 4 degrees of freedom, mean 0, and the same Mat\'{e}rn covariance function used in section \ref{linear_sim}. For this scenario, we use estimated imputation parameters and follow all other default simulation settings outlined in section \ref{est_sim}, performing the simulations 100 times. For brevity, we leave $N$ and $J$ fixed and only explore differing values of $m$. Table \ref{tab:ttab_m} summarizes the results of these simulations.

Upon inspection, the use of a multivariate t distribution as opposed to a Gaussian process to generate the data yields little difference in terms of estimation performance across all four methods. The MISE results of MuC shown in table \ref{tab:ttab_m} are strikingly similar to those shown in \ref{tab:estm} and the performance of MuC relative to the other three imputation approaches seems unchanged as well. This indicates some level of robustness among all four approaches to the assumption of Gaussian data.

\begin{table}[h]
\centering
\scalebox{1}{
\begin{tabular}{lllllllllllll}
  \hline
   m & \multicolumn{4}{c}{w = 0} & \multicolumn{4}{c}{w = 5} & \multicolumn{4}{c}{w = 10} \\ \cmidrule(lr){2-5}\cmidrule(lr){6-9}\cmidrule(lr){10-13}
  & MeC & MuC & MeU & MuU & MeC & MuC & MeU & MuU & MeC & MuC & MeU & MuU \\
 \hline
2 & 0.46 & 0.04 & 0.05 & 0.01 & 5.35 & 5.16 & 9.56 & 7.29 & 18.98 & 18.88 & 38.49 & 28.96 \\ 
  5 & 0.04 & 0.02 & 0.02 & 0.01 & 4.31 & 4.00 & 5.38 & 5.17 & 14.26 & 13.85 & 21.50 & 20.61 \\ 
  10 & 0.02 & 0.01 & 0.01 & 0.01 & 3.76 & 3.60 & 4.85 & 4.79 & 13.38 & 13.38 & 19.38 & 19.18 \\ 
  20 & 0.01 & 0.01 & 0.01 & 0.01 & 3.51 & 3.44 & 4.47 & 4.47 & 13.12 & 13.11 & 17.84 & 17.76 \\ 
   \hline
\end{tabular}
}
\caption{MISE as m increases, using estimated parameters and sampling the functional covariate from a multivariate t distribution} 
\label{tab:ttab_m}
\end{table}

\subsection{Logistic Regression}\label{logistic_sim}

Finally, since our approach can easily be applied to logistic scalar-on-function regression as well as the simple linear model, we compare performance of the four approaches in a logistic regression setting as well. For brevity, we omit the simulations using the true simulation parameters and only provide results for simulations using estimated parameters.

We begin by simulating $Y_i\overset{iid}{\sim}Berrn(p)$, for $i=1,\cdots,N$ with $p=0.5$. Then we simulate $X_i|Y_i=y \overset{iid}{\sim} N(\mu_y,C_X(s,t))$ evaluated on an equally-spaced length-100 grid in $[0,1]$ where $\mu_0(t)\equiv 0$, $\mu_1(t)=v_1(t)+v_2(t)$, and  $C_X(t,s)$ is the same Mat\'{e}rn covariance given in equation \ref{eq:matern}. We again add noise to the $X_i$ using $\sigma_\delta^2 = 0.5$, and randomly sample $m$ of the 100 grid points to observe for each curve. According to Theorem \ref{t:log}, since our choices of $\mu_0$ and $\mu_1$ satisfy $||C^{-1}_X(\mu_1 - \mu_0)||^2_{L_2} < \infty$, we use $\beta(t)=C^{-1}_X(\mu_1 - \mu_0)=\frac{v_1(t)}{\lambda_1} + \frac{v_2(t)}{\lambda_2}$. Estimation of imputation parameters is performed as in section \ref{est_sim}, and a logistic regression model is fit using the first $J$ FPCs.

For these simulations, we choose $N=400$, $m=2$, and $J=2$ by default and show results for $N \in \{ 100, 200, 400, 800 \}$, $m \in \{ 2, 5, 10, 20\}$, and $J \in \{ 1\cdots,6\}$, with 100 runs for each setting. Note that we do not adjust the signal in these simulations as we did in the simulations for the linear model. This is partially due to brevity and partially due to the inherent instability of logistic regression. We found that simulation results were particularly sensitive to the signal such that when the signal was too large and near-perfect classification could be achieved, all methods performed quite poorly and comparisons were less interesting. Likewise, we increase the default sample size to $N=400$ for these simulations to insert additional stability. As before, we report MISE for the estimates resulting from each of the four imputation methods.

As in the previous simulations, we first fixed $m$ and $N$ and considered the MISE for different values of $J$. The results are shown in table \ref{tab:logisticJ}. Clearly $J = 2$ results in the lowest MISE for all methods except MeC. This resulted from MeC's erratic behavior, which was best ameliorated by retaining only the first FPC, essentially imposing some smoothness on MeC by forcing it to produce a linear estimate. Conversely, MuU is noteworthy due to how resolutely smooth its estimates remained regardless of the choice of $J$. 

\begin{table}[h]
\centering
\begin{tabular}{lllll}
  \hline
  J & MeC & MuC & MeU & MuU \\
 \hline
1 & 22.82 & 22.00 & 17.08 & 19.20 \\ 
  2 & 316.16 & 4.08 & 2.52 & 14.02 \\ 
  3 & 173160.16 & 110.44 & 70.59 & 14.09 \\ 
  4 & 938930.00 & 3037.01 & 3661.15 & 14.70 \\ 
  5 & 1880976.95 & 399511.49 & 83670.09 & 17.15 \\ 
  6 & 5339327.24 & 1816022.46 & 1706713.88 & 36.26 \\ 
   \hline
\end{tabular}
\caption{MISE of $\hat{\beta}(t)$ as $J$ increases, 
                          for a logistic model with estimated imputation parameters.} 
\label{tab:logisticJ}
\end{table}
\begin{table}[h]
\centering
\begin{tabular}{lllll}
  \hline
  N & MeC & MuC & MeU & MuU \\
 \hline
100 & 3212.76 & 20.46 & 6.65 & 14.12 \\ 
  200 & 1973.24 & 11.96 & 4.51 & 14.03 \\ 
  400 & 371.82 & 4.08 & 2.52 & 14.03 \\ 
  800 & 171.69 & 2.79 & 2.26 & 14.02 \\ 
   \hline
\end{tabular}
\caption{MISE of $\hat{\beta}(t)$ as $N$ increases, 
                          for a logistic model with estimated imputation parameters.} 
\label{tab:logsticN}
\end{table}
Table \ref{tab:logsticN} shows the MISE as $N$ increases from 100 to 800. Unlike with the linear model, MeU is more accurate than MuC for a small sample size of 100, but as $N$ grows, MuC quickly catches up and is comparable to MeU. It is encouraging that despite observing each curve only twice, for a sample size of 800, MuC performs quite well. The behavior of both MuU and MeC in these simulations is also quite interesting. Increasing $N$ does little to improve the accuracy of MuU, which tends to drastically oversmooth its estimates, and while MeC improves as $N$ grows, its instability renders it incomparably inaccurate even for a larger sample size. 

\begin{table}[h]
\centering
\begin{tabular}{lllll}
  \hline
  m & MeC & MuC & MeU & MuU \\
 \hline
2 & 371.82 & 4.08 & 2.53 & 14.02 \\ 
  5 & 18.86 & 0.73 & 1.09 & 5.83 \\ 
  10 & 3.37 & 0.46 & 0.35 & 1.97 \\ 
  20 & 0.85 & 0.28 & 0.17 & 0.54 \\ 
   \hline
\end{tabular}
\caption{MISE of $\hat{\beta}(t)$ as $m$ increases, 
                          for a logistic model with estimated imputation parameters.} 
\label{tab:logisticm}
\end{table}
Table \ref{tab:logisticm} tells a similar but distinct story. First of all, each method improves as $m$ increases, and the improvements in accuracy are more stark for small changes in $m$ than they were for small changes in $N$. This is most evident in the improvement of MeC and MuU, which are uncompetitve when $m = 2$, but approach MuC and MeU in accuracy as $m$ grows to 20. MuC and MeU are quite comparable across the different values of $m$, though MeU boasts a slight advantage when $m=2$. 

\section{Application to Macrocephaly}\label{sec:app}

Equipped with a new approach to scalar-on-function regression for sparsely and irregularly sampled functional data, we are prepared to revisit the macrocephaly data. One important scientific question that we hope to answer is whether and how children's head circumference trajectories are associated with their chance of having a pathology related to macrocephaly. 
We follow the approach outlined in Section \ref{sec:logistic} for imputing the curves in a logistic scalar-on-function regression context. Due to the association between height and head circumference, we use a ratio of head circumference to height instead of raw head circumference. Thus, according to model \ref{eq:log}, $Y_i$ indicates the presence or absence of pathology, $p_i$ the probability of the pathology occurring, and $X_i(t)$ the height-adjusted head circumference at age $t$ of the $i^{th}$ subject.

Due to the prohibitive size of the data set compounded with the rarity of the pathology in the data, we fit a logistic regression model using a subset of $N = 800$ subjects. Specifically, we retained all 85 cases in our subsample, and the remaining 715 controls were randomly selected according to a stratified sampling scheme to roughly match the distribution of sampling frequency between cases and controls. The subjects in the resulting subsample had an average of 6.5 measurements. For the sake of comparison, we imputed the curves $\{ X_i(t): i =1\cdots,800\}$ according to all four imputation approaches outlined in Section \ref{sec:sims}. For the multiple imputation approaches, $K = 10$ imputations were used. The imputed curves are depicted below in figure \ref{fig:impcurves}. It is clear that major differences exist between the conditional and unconditional approaches, where, for example, the imputations are noticeably different towards the end of the age range.

\begin{figure}[h]
   \centering
   \includegraphics[scale=.4]{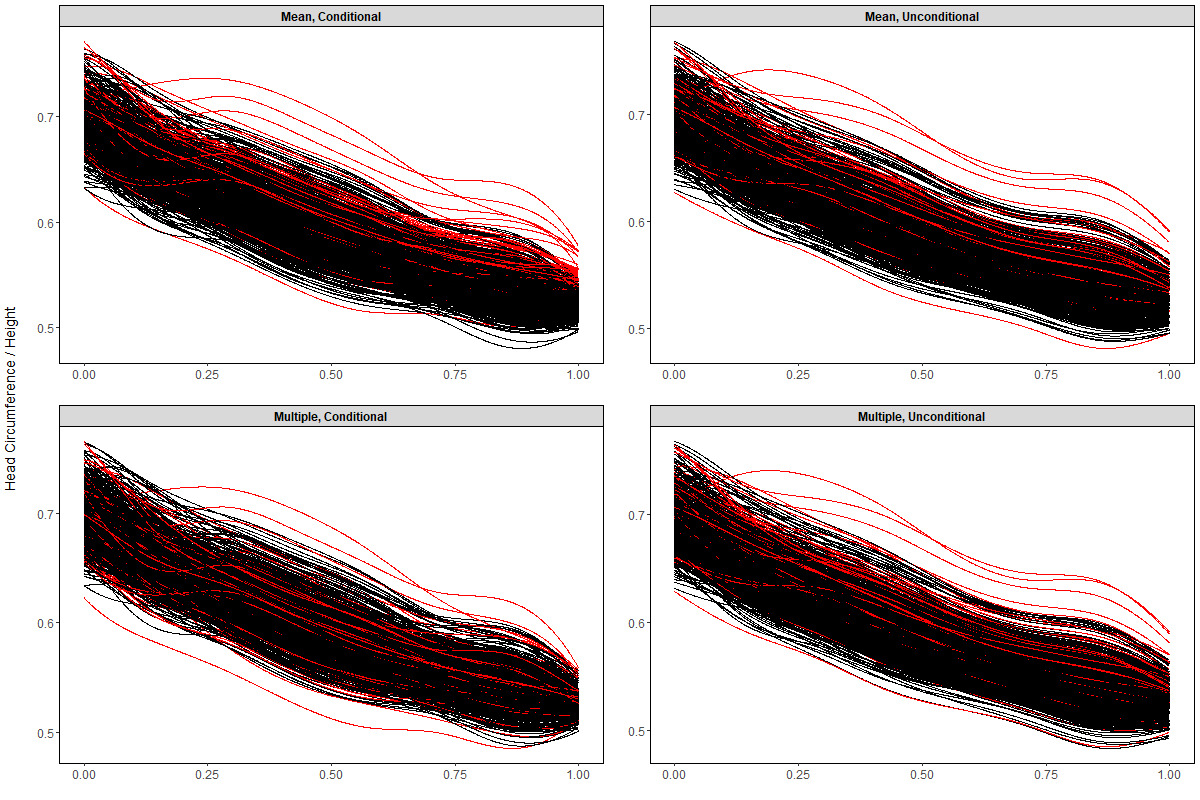}
   \caption{Imputed curves for each of the four imputation approaches. Red lines represent cases, black lines represent controls.}
   \label{fig:impcurves}
\end{figure}

\begin{figure}[h]
   \centering
   \includegraphics[scale=.5]{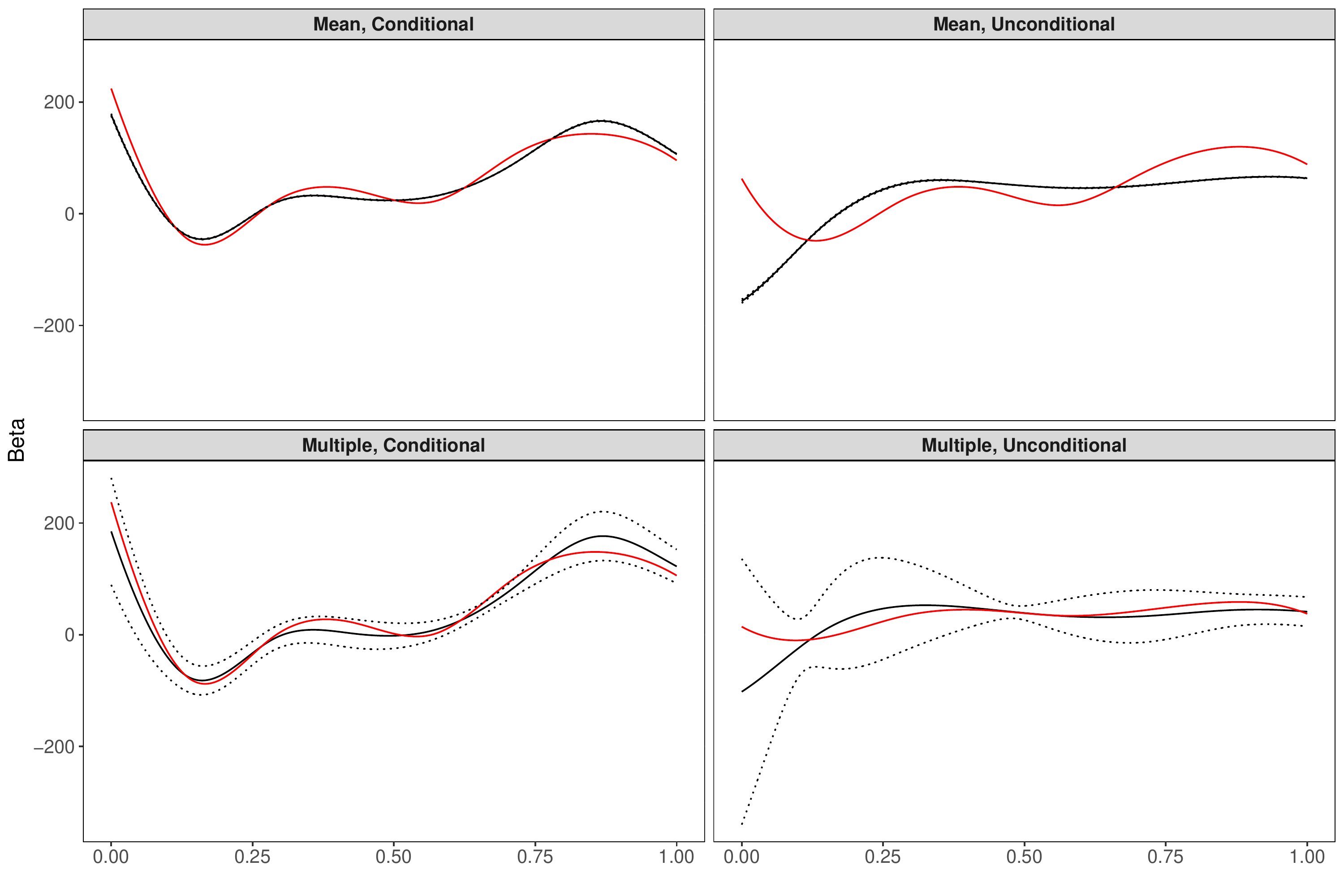}
   \caption{Estimated coefficient functions. Solid black lines are the estimated coefficient functions, dotted black lines are the associated 95\% pointwise confidence bands. Red lines are the estimated coefficients from the resampled data.}
   \label{fig:allbetas}
\end{figure}

After imputing the height-adjusted head circumferences, a logistic regression model was subsequently fit for each of the four approaches. The estimated coefficient functions are shown in black in figure \ref{fig:allbetas}, along with their 95\% pointwise confidence intervals (the black dotted lines). There is reasonable similarity in the coefficient functions estimated by the two unconditional imputation approaches and even more agreement between the estimated coefficient functions from the conditional imputation approaches. P-values for the test of a non-zero effect are presented in table \ref{tab:pvals}. As was pointed out earlier, it is wise to be skeptical of the p-values produced from mean imputation as they do not properly incorporate imputation-specific uncertainty. 
However, in this case, the conclusion one would draw from the MuC approach is consistent with the mean imputation approaches as the p-value is still sufficiently low to reject the null hypothesis at any reasonable level of significance. The only p-value which is not convincingly low is that of MuU which, as we saw in the simulations, has lower power than the other approaches and is prone to miss signals.

\begin{table}[h!]
\centering
\begin{tabular}{ |c|c|c|c| }
 \hline
 MeC & MeU & MuU & MuC\\
 \hline
 0.000000 & 0.000000 & 0.122862 & 0.000000\\
 \hline
\end{tabular}
\caption{P-values for test of no effect according to all four imputation approaches applied to the macrocephaly data.}
\label{tab:pvals}
\end{table}

One of the main benefits of using a multiple imputation approach is the ability to better estimate uncertainty due to imputation. This is evident in the much wider confidence bands for the multiple imputation approaches compared to their mean imputation counterparts (the confidence bands for the mean imputation approaches are so narrow that they are barely visible). To gauge the suitability of the confidence bands we sampled with replacement from the subsample, and duplicated the imputation and estimation steps again (basically a one sample bootstrap). The resulting estimates are represented by the red lines. For the mean imputation approaches to have accurate confidence bands would require the new estimates to be nearly identical to the original estimates, which is clearly not the case. The resampled estimates for the multiple imputation approaches, conversely, are well captured by the confidence bands.

\begin{figure}[h]
  \centering
  \includegraphics[scale = .5]{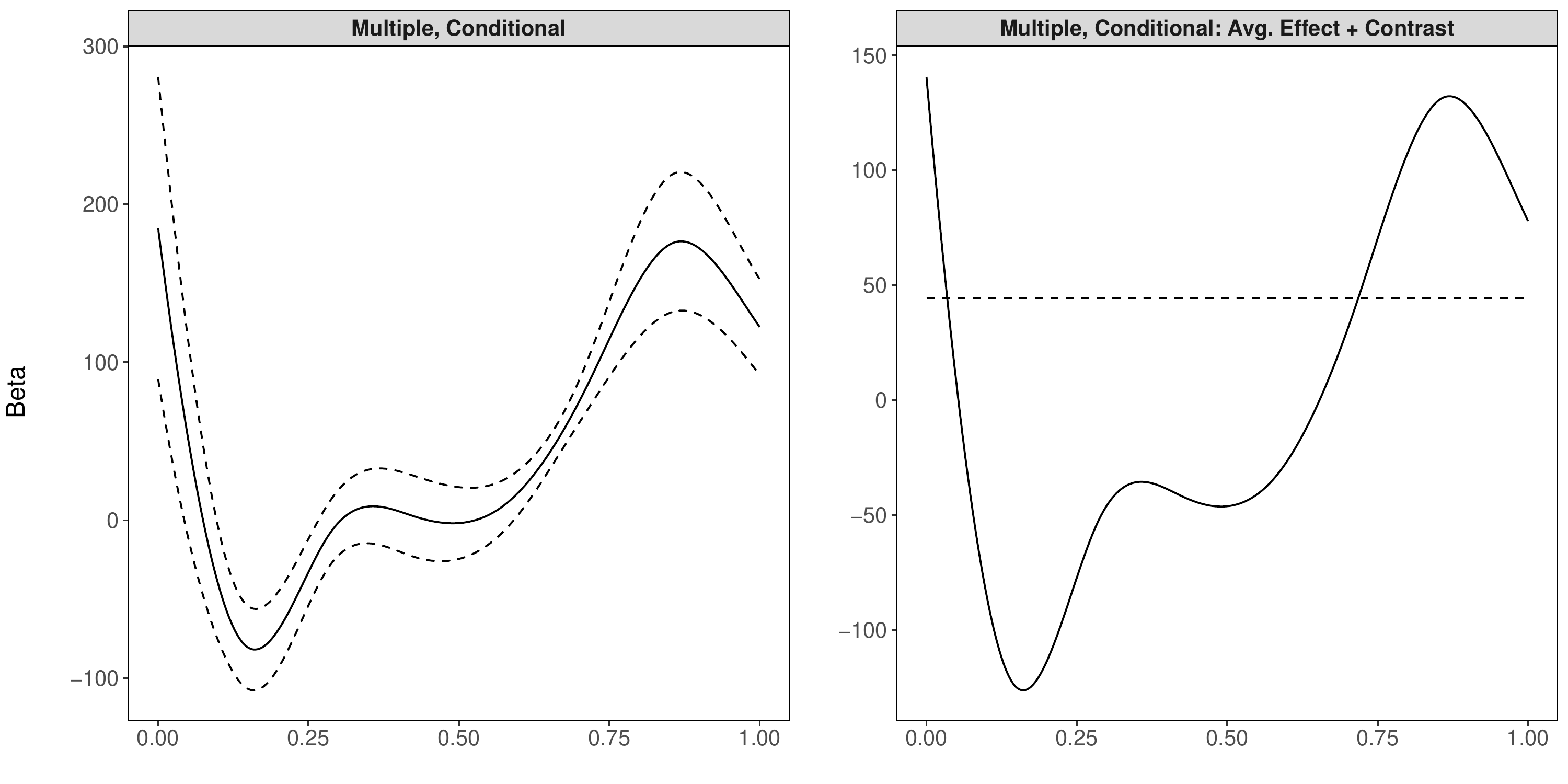}
  \caption{ Left panel: Estimated coefficient function using the MuC approach (solid line) and associated 95\% pointwise confidence bands (dashed lines). Right panel: Average effect (dashed line) and estimated coefficient function less the average (solid line).}
  \label{fig:muc_beta}
\end{figure}

Focusing solely on the MuC approach, the left panel of figure \ref{fig:muc_beta} shows a zoomed-in view of the estimated coefficient function. 
Notice the swift decline from large positive values to negative values early in the domain, followed by a fairly steady increase back to positive values until the end of the domain. Though it is tempting to conclude that the association between the ratio of head circumference to height and the probability of developing a pathology vacillates from largely positive, to negative, and back to positive, 
this misses the fact that the effect must be interpreted jointly over the entire domain. 

To aid our interpretation, we turn to the right panel of figure \ref{fig:muc_beta}, which decomposes the coefficient function into two parts: the average effect (which is constant over time) and the total effect less its average. The latter, depicted by the solid line, suggests a contrast between the negative values occurring in much of the first half of the domain, and the positive values occurring in the second half of the domain. Such a contrast can be viewed as an indication that the velocity of head circumference-to-height growth drives the distinction between the groups. Taken together then, two components imply an elevated risk of developing the pathology: larger average head circumferences and unusual head circumference growth rates (both relative to height).

\section{Asymptotic Theory}\label{sec:theory}
In this section we provide a preliminary asymptotic justification for our procedure.  This theory is by no means complete as our focus is more methodological.  However, it establishes  consistency of the MISFIT method in this particular setting; note that carrying out PACE to impute the curves/scores would not result in a consistent estimate of $\beta(t)$, as discussed in Section \ref{sub:lm}.  This theory focuses on FPCA estimation for the linear model and for the case where the number of points per curve is the same and fixed $m_i \equiv m$, though there are now a variety of additional settings one can consider.  For ease of exposition, we clearly list all of the assumptions below, even those that have already been discussed.  

\begin{assumption} \label{a:main}
We make the following modeling and estimation assumptions.  
\begin{enumerate}
\item The predictors, $X_i(t) \sim \mcN(\mu_X, C_X)$, are iid Gaussian processes and are independent of the iid errors, $\vep_i \sim \mcN(0, \sigma^2_\vep)$.
\item The outcome, $Y_i$, is given by 
\[
Y_i =  \alpha + \int \beta(t) X_i(t) + \vep_i,
\]
where $\alpha$ and $\beta(t)$ are deterministic parameters.
\item The outcomes are observed, but the predictors are only observed at points $t_{ij}$ and with error: $x_{ij} = X_i(t_{ij}) + \delta_{ij}$.  The errors $\delta_{ij}$ are iid normal random variables with mean 0 and variance $\sigma^2_\delta>0$.  They are also independent of all other quantities.
\item  The function $\beta(t)$ lies in the span of the first $p$ eigenfunctions of $C_X$. 
\item  We have consistent estimates of the parameters, $\mu_X$, $C_X$, $\sigma_\delta^2$, $C_{XY}$, $\mu_Y$, and $\sigma_Y^2$ in the sense that
\begin{align*}
& \sup_t |\hat \mu_X(t) - \mu_X(t)| = o_P(1) \quad
\sup_{t,s} | \hat C_X(t,s) - C(t,s) | = o_P(1) \quad 
\sup_t |\hat C_{XY}(t) - C_{XY}(t)| = o_P(1)  \\
& | \hat \sigma_\delta - \sigma_\delta| = o_P(1) \quad
| \hat \mu_Y - \mu_Y| = o_P(1) \quad 
| \hat \sigma_Y^2 - \sigma_Y^2| =o_P(1).
\end{align*}
\item The first $p+1$ eigenvalues of $C_X$ are distinct.
\item The number of points per curve is fixed and the same for every $i$, i.e. $m_{i} \equiv m >0$.
\end{enumerate}
\end{assumption}
We now discuss each of these assumptions in more detail.  The first three assumptions are simply our modeling assumptions.  The fourth assumption makes the asymptotics easier to derive by assuming that there is no truncation error when expressing $\beta(t)$ using the first $p$ eigenfunctions of $C_X$.  While convenient, it is clearly not true in general and at best an approximation.  One can see \citet{cai2012minimax} for discussion on the interplay between $\beta(t)$ and the eigenfunctions of $C_X$.  The fifth assumption guarantees that we have consistent estimates of the various parameters needed for imputation.  These can be computed using any number of methods (splines, local smoothing, etc.) and such an assumption allows us to avoid assuming a specific approach (as well as listing all of the additional technical assumptions each approach would require).  The sixth assumption is common in FDA and guarantees that the eigenfunctions can be consistently estimated, though this assumption can be relaxed \citep{petrovich2017asymptotic}.  The last assumption is again for convenience, providing simpler expressions and asymptotic computations.  However, this assumption can be relaxed, and there is a great deal of interest in the interplay between convergence rates and assumptions on $m$ (as well as $\beta(t)$).  As our goal is methodological, we do not explore these dynamics in the present work.

Define the estimated quantities
\[
\hat \bmu_i = \hat \bA_i \hat \bB_i \bd_i \qquad  \hat \Sigma_i =  \hat \blambda_p - \hat \bA_i ^\top \hat \bB_i \hat \bA_i.
\]
We then draw $\hat \bxi_i$ from a multivariate normal distribution with the above mean and variance respectively.  More specifically, let $\bZ_i$ be iid standard normal random vectors (i.e. mean zero and identity covariance). Then, without loss of generality we assume $\hat \bxi_i$ is generated as
\[
\hat \bxi_i = \hat \bmu_i + \hat \Sigma_i^{1/2} \bZ_i.
\]
Here we take $\Sigma^{1/2}$ to be the symmetric positive definite square-root so that it is unique.  We then have the following result.  
\begin{theorem} \label{t:main}
Under Assumption \ref{a:main} we have that
\[
\sup_t| \hat \beta(t) -  \beta(t)| = o_P(1),  
\] 
as $N \to \infty$.
\end{theorem}

\section{Conclusions}\label{sec:con}

In this paper, we have introduced a multiple imputation approach, MISFIT, to performing scalar-on-function linear regression in the presence of sparse and irregular functional data. This approach yields consistent estimates and captures the variation due to imputing the functional covariates, thus enabling more reliable statistical inference. We showed that this method can be applied more broadly to other functional generalized linear models by utilizing an appropriate imputation model, and in particular demonstrated its use in a logistic regression setting. Extensive simulations also illustrated the value of MISFIT over existing methods for fitting scalar-on-function regression models when the functional covariate is irregularly and sparsely sampled.

Beyond the general effectiveness of MISFIT, one other takeaway is worth highlighting from the simulations: the accuracy of MISFIT is quite sensitive to the estimates of the imputation parameters, especially in highly sparse and irregular designs. In practice, there are multiple justifiable approaches to estimate these parameters and our results are based on the use of {\tt fcr}, which employs a spline-based approach. However, we did see some preliminary evidence in favor of the local polynomial smoothing approach used by {\tt fdapace}. These differences were not explored in depth here, but could warrant further investigation given their practical implications.

Other future work will involve extending this multiple imputation approach to more complicated models. It would be beneficial, for instance, to apply the same approach to GAMs. While the same ideas can be carried over easily enough, establishing a compatible imputation model could prove challenging. Even including multiple functional covariates in the linear or logistic models requires more careful thought than one would hope for such a seemingly direct extension. Much additional work is also left surrounding the asymptotic theory. While consistency of the estimated coefficient functions was established, it was done so under fairly strong assumptions. The next useful result would be to establish minimax rates of convergence and, in particular, to investigate the relationship between the number of observed points per curve and these rates.

\bibliographystyle{chicago}
\setlength{\bibsep}{0pt plus 0.3ex}
\bibliography{pace}

\clearpage


\appendix 
\begin{center}
\title{{\huge Supplementary Material: Highly Irregular Functional Generalized Linear Regression with Electronic Health Records}}	
\end{center}

\begin{center}
	\author{{\large Justin Petrovich\qquad \and Matthew Reimherr\qquad \and Carrie Daymont}}
\end{center}

\section{Some Technical Notes on Cameron-Martin Theory}
In this section, we provide some additional details for readers interested in the background on equivalence/orthogonality of Gaussian measures over Hilbert spaces.  All details can be found in \citet{bogachev1998gaussian}, though we have attempted to formulate the ideas in a way that is more natural for an FDA audience.  Recall that we assume $X_i \in L^2[0,1]$ almost surely, meaning, the sample paths of $X_i$ are square integrable with probability 1.  The space $L^2[0,1]$ is a Hilbert space with inner product
\[
\langle f, g \rangle = \int_0^1 f(t) g(t) \ dt.
\]
Let $Z \sim \mcN(0, C)$, with $Z \in L^2[0,1]$ and assume $C$ has full rank.  Then we can use $Z$, or equivalently $C$, to define a second inner product, which we denote as
\[
\langle f, g \rangle_{\mcK} = \E[\langle f, Z \rangle \langle g,Z \rangle] = \int \int f(t) C(t,s) g(s) \ dt \ ds.
\]
The resulting quantity above can also be viewed as $\langle f, C g\rangle$, $\langle C f, g \rangle$, or $\langle C^{1/2} f, C^{1/2} g \rangle$, where $C^{1/2}$ is the symmetric square-root.  Since $Z \in L^2[0,1]$, one can show that $C$ and $C^{1/2}$ are well defined continuous linear operators over $L^2[0,1]$.  As long as $C$ has full rank, then this $\langle \cdot, \cdot \rangle_{\mcK}$ a valid inner product over $L^2[0,1]$.  If $C$ were to have a nontrivial null space then one can ``cut out'' those dimensions from $L^2[0,1]$, though for ease of exposition we don't pursue that here.

Since $Z \in L^2[0,1]$, one can show that $\langle f, g \rangle_{\mcK} < \infty$ for any $f,g\in L^2[0,1]$, however, $L^2[0,1]$ is not complete under this inner product.  The completion of $L^2[0,1]$ under $\langle \cdot, \cdot \rangle_{\mcK}$ we call $\mcK$.  Typically, the elements of $\mcK$ are much ``rougher'' than those in $L^2[0,1]$, since they can rely on $C$ to ``smooth them out''.  While $\mcK$ is much larger than $L^2[0,1]$, we now take a smaller subset, denoted as $\mcH \subset L^2[0,1]$.  In particular, each element $h \in L^2[0,1]$ defines a continuous linear functional over $L^2[0,1]$ through $f \to \langle h, f \rangle$.  This functional naturally extends to one over $\mcK$, but it no longer need be continuous since $\mcK$ is much larger.  So we say $h \in \mcH$ if the corresponding functional is continuous in $\mcK$, meaning $\langle h , f \rangle$ is well defined and bounded for all $\|f \|_{\mcK} \leq 1$.  However, since $\mcK$ is a Hilbert space, this implies that there exists an element $f_h \in \mcK$ such that $\langle f_h , g \rangle_{\mcK} = \langle h, g \rangle$ for all $g \in \mcK$.

The space $\mcH$ is called the {\it Cameron-Martin space} of $C$ (or equivalently $Z$) and is isomorphic to the reproducing kernel Hilbert space generated by $C$, though one needs to take slight care since $L^2[0,1]$ consists of equivalence classes.  It is also a Hilbert space in its own right under the inner product $\langle h_1, h_2 \rangle_{\mcH} := \langle f_{h_1}, f_{h_2} \rangle_{\mcK} $, where $f_{h_i}$ is the corresponding element of $\mcK$ for $h_i$.  Less abstractly, one has the relationship $C(f_{h_i}) = h_i$.  This also provides a less abstract formulation of $\langle \cdot, \cdot \rangle_{\mcH}$ as one can equate the following inner products
\[
\langle h_1, h_2 \rangle_{\mcH} = \langle C^{-1} h_1, C^{-1} h_2 \rangle_{\mcK} = \langle h_1, C^{-1} h_2 \rangle,
\]
which can be easier to interpret.  One benefit of considering a more abstract formulation is the ability to generalize beyond Hilbert spaces by making judicious use of dual spaces.  In particular, one can work with in any locally convex vector space, which includes Hilbert spaces, Banach spaces, and Frechet spaces.

The Cameron-Martin space of a Gaussian measure completely characterizes the equivalence/orthogonality of Gaussian processes that are translations of the original.  In particular, two Gaussian measures $P_1 \sim \mcN(\mu_1,C)$ and $P_2 \sim \mcN(\mu_2,C)$ are equivalent if $\Delta= \mu_2 -\mu_1$ resides in the Cameron-Martin space of $C$, i.e., the RKHS generated by $C$, and orthogonal otherwise.  Furthermore, it provides a recipe for defining the density of $P_2$ with respect to $P_1$.  This is critical in infinite dimensional spaces as one does not have a Lebesgue measure to rely on when defining densities.  If we relabel $\mu_1 = \mu$ and $\mu_2 = \mu + \Delta$, then the density (Radon-Nikodym derivative) of $P_2$ with respect to $P_1$ is
\[
\frac{dP_2}{dP_1} (x) = \exp\left\{ \langle x - \mu, \Delta \rangle_{\mcH}  - (1/2) \| \Delta\|_{\mcH} \right\}.
\]
While $\langle x - \mu, \Delta \rangle_{\mcH}$ is well defined and finite almost surely when $x$ is a sample path of $P_1$, it need not be for an arbitrary element of $L^2[0,1]$.  However, if in addition $\| C^{-1} \Delta\| < \infty$ (equivalently, $\Delta$ resides in the Cameron-Martin space of $C^2$), then $C^{-1} \Delta \in L^2[0,1]$, and one can write
\[
\langle x - \mu, \Delta \rangle_{\mcH} 
= \int (x(t) - \mu(t)) \beta(t) \ dt,
\]
where $\beta= C^{-1} \Delta$.  If one only has that $\|C^{-1/2} \Delta\| < \infty$ (i.e. $\Delta$ resides in the Cameron space of $C$, but not $C^2$), then $\beta$ cannot be identified as an element of $L^2[0,1]$.

\section{Proof of Theorem \ref{t:log}}
Let $X_i | Y_i=0 \sim \mcN(\mu, C_X):=P_0$ and $X_i | Y_i=1 \sim \mcN(\mu + \Delta,C_X):=P_0$.  Let $\mcH$ denote reproducing kernel Hilbert space of $C_X(t,s)$, equivalently, it is the Cameron-Martin space of the Gaussian process $\mcN(0,C_X)$.   
From 
Corollary 2.4.3 of \citet{bogachev1998gaussian}, 
 $P_0$ and $P_1$ are equivalent if $\|C^{-1/2}\Delta \| < \infty$ and orthogonal otherwise.  Thus, this quantity must be finite for there to exist a nondegenerate logistic regression model.  To find the form of $\alpha$ and $\beta$ in the logistic regression model, we can use the equivalence of the two measures to  which has the closed form expression \citep{mirshani2017existence,rao1963discrimination} 
\[
\frac{d P_1}{ dP_0}(x) = \exp\left\{  
T_\beta(x - \mu)
- (1/2)\|\Delta \|^2_{\mcH} \right\},
\]
where $T_\beta$ is a linear functional, however, it is not continuous over all of $L^2[0,1]$ unless $\| C_X^{-1} \Delta\| < \infty$, in which case it can be written as
\[
T_\beta(x - \mu) = \langle x - \mu, C^{-1} \Delta \rangle.
\]

We can compute the logit of $p_i$ as
\begin{align*}
\logit(p_i) = \log \left(  \frac{P(Y_i = 1 | X_i = x_i)}{P(Y_i=0 | X_i=x_i)} \right)
 & = \log \left( \frac{d P_1}{ dP_0}(x_i) \left[ \frac{P(Y_i=1)}{P(Y_i=0)}\right] \right) \\
& = T_\beta(x - \mu)
- (1/2)\|\Delta \|^2_{\mcH}
+ \log \left( \frac{P(Y_i=1)}{P(Y_i=0)} \right).
\end{align*}
Thus, we see that, in the context of logistic regression, if $\|C_X^{-1}\Delta\| < \infty$ then
\[
\beta = C_X^{-1} \Delta  
\qquad
\alpha = - \langle \mu, C_X^{-1} \Delta \rangle- (1/2)\|\Delta \|^2_{\mcH}+ \log \left( \frac{P(Y_i=1)}{P(Y_i=0)} \right),
\]
as claimed.

\section{Proof of Theorem \ref{t:glm}}
Suppose that $X | Y=y \sim \mcN( \mu + T(y) \Delta , C_X ) $ and that $\| C_X^{-1/2} \Delta \|$ is finite.  Then $X|Y=y$ has a density with respect to $\mcN(\mu,C_X)$ \citep{bogachev1998gaussian}:
\[
f_{X|Y}(x|y) = \exp\left\{T(y) \langle x - \mu, C_X^{-1} \Delta \rangle  - (1/2) T(y)^2 \| C_X^{1/2} \Delta\|^2\right\}.
\]
Bayes' rule implies that $Y|X$ has density
\begin{align*}
f_{Y|X}(y|x) & \propto f(x|y) f(y)  \\
& = \exp\left\{T(y) \langle x - \mu, C_X^{-1} \Delta \rangle  - (1/2) T(y)^2 \| C_X^{-1/2}\Delta\|^2 )\right\} f(y) \\
& = \exp\left\{T(y) \langle x - \mu, C_X^{-1} \Delta \rangle \right\} h(y),
\end{align*}
where $h(y) = \exp\{- (1/2) T(y)^2 \| C_X^{-1/2}\Delta\|^2 \} f(y) $.  The normalizing constant is given by
\[
\exp\{A(\eta_x)\}= \int h(y) \exp\left\{T(y) \langle x - \mu, C_X^{-1} \Delta \rangle \right\}  \ dy.
\]
Which now gives exactly the correct form where $\eta_x = \langle x - \mu, C_X^{-1} \Delta \rangle$. This implies that $\beta = C_X^{-1}\Delta$ 
as claimed.  If $\|C_X^{-1} \Delta\|$ is not finite, then $\beta$ is not an element of $L^2[0,1]$, but $\langle x - \mu,  C_X^{-1}\Delta \rangle$ is still well defined, almost surely, when $x$ is a realization of $X$.  If $\|C_X^{-1/2} \Delta\|$ is not finite then the collection of distributions $\{X|Y=y\}$ are all mutually singular, meaning $Y$ can be perfectly predicted from $X$ with probability one and thus $Y|X$ is degenerate.

\section{Proof of Theorem \ref{t:main}}
\begin{proof}
Recall that
\[
\hat \beta(t) = \sum_{j=1}^p \sum_{i=1}^N \frac{\hat \xi_{ij} Y_i}{\hat \lambda_j N } \hat v_j(t)
\qquad \text{and} \qquad
\beta(t) = \sum_{j=1}^p \sum_{i=1}^N \frac{\E[ \xi_{1j} Y_1]}{\lambda_j} v_j(t).
\]
We establish our theorem via Slutsky's lemma, showing that each term in $\hat \beta(t)$ converges to its desired population counterpart.  Trivially, one has that
$\sup_t | \hat v_j(t) - v_j(t) | =o_P(1)$ and $|\hat \lambda_i - \lambda_i| = o_P(1)$ since the estimated covariance is convergent and the population eigenvalues are distinct.  Therefore we need only show that
\[
\hat \bzeta: = \sum_{i=1}^N \frac{\hat \bxi_{i} Y_i }{N} \to \E[\bxi_{i} Y_i] =: \bzeta.
\]
To do this, we decompose $\hat \bzeta$ as
\[
\hat \bzeta = \bT_1 + \bT_2 + \bT_3 + \bT_4,
\]
where
\begin{align*}
\bT_1 := \sum  \frac{\bmu_i Y_i }{N} \quad 
\bT_2 := \sum  \frac{\Sigma_i^{1/2} \bZ_i Y_i}{N} \quad 
\bT_3 := \sum  \frac{(\hat \bmu_i - \bmu_i) Y_i}{N} \quad
\bT_4 := \sum  \frac{(\hat \Sigma_i^{1/2} - \Sigma_i^{1/2}) \bZ_i Y_i}{N}.
\end{align*}
We will show that the first term converges to $\bzeta$ and the others to zero.  

For $\bT_1$, one has that it converges to $\bzeta$ using properties of conditional expectation and the SLLN.  For the second term, it is a sum of mean zero independent random vectors, thus the entire sum converges to zero again by the SLLN.  For the third term we have, by Lemma \ref{lem1}
\[
\left| \sum  \frac{(\hat \bmu_i - \bmu_i) Y_i}{N} \right|
\leq \sup_i |\hat \bA_i \hat \bB_i - \bA_i \bB_i|  \sum \frac{|\bd_i Y_i|}{N} = o_P(1) O_P(1) = o_P(1).
\]
Lastly, turning to the fourth term, we have again by Lemma \ref{lem1}
\[
\left|
\sum  \frac{(\hat \Sigma_i^{1/2} - \Sigma_i^{1/2}) \bZ_i Y_i}{N}
\right|
\leq \sup |\hat \Sigma_i^{1/2} - \Sigma_i^{1/2}| \sum \frac{|\bZ_i Y_i|}{N}
= o_P(1) O_P(1),
\]
which completes the proof.

\end{proof}

\begin{lemma}\label{lem1}
Under Assumption \eqref{a:main} we have
\begin{align}
\sup_i | \bA_i - \hat \bA_i| = o_P(1)  \label{e:a} \\
\sup_{i} | \bB_i - \hat \bB_i| = o_P(1) \label{e:B} \\
\sup_i | \bA_i \bB_i - \hat \bA_i \hat \bB_i| = o_P(1) \label{e:ab}  \\
\sup_i | \Sigma_i^{1/2} -  \hat \Sigma_i^{1/2}| = o_P(1) \label{e:sig}
\end{align}
\end{lemma}
\begin{proof} \ \\
It follows that \eqref{e:a} holds since $| \hat v_j(t_{ij}) - v_j(t_{ij})| \leq \sup_t |v_j(t) -\hat v_j(t)| \leq C \sup_{t,s} |C(t,s) - \hat C(t,s)|$, and both $m$ and $p$ are finite (reall $\bA_i$ is an $m \times p$ matrix). \ \\

For \eqref{e:B} notice that $\sigma_Y = \sigma_\vep^2 + \int \beta(t) C_X(t,s) \beta(s) \ ds$ and $C_{XY}(t) = \int C(t,s) \beta(s) \ ds$.  Thus, $\bB_i^{-1}$ is the sum of two positive definite matrices, one of which is diagonal ($diag\{\sigma_\vep^2, \sigma_\delta^2, \dots\}$), and the same holds for $\hat \bB_i$. Since the $\bB_i^{-1}$ only differ at which time points are observed, we can construct a closed convex set, $E$, in the cone of positive definite matrices (in the space of symmetric matrices) that includes all of the $\bB_i^{-1}$ for any $N$.  
Since $\hat \bB_i^{-1}$ converges to $\bB_i^{-1}$ uniformly in $i$, it follows that, for $N$ large, the set $E$ can be increased by a small $\epsilon>0$ (i.e. include any matrix within $\epsilon$ of a point in $E$) to include all of the $\hat\bB_i^{-1}$ and still only include positive definite matrices.   On this new set the inverse map is continuously differentiable, and so we can find a constant, $C$, which does not depend on $i$ such that, for $N$ large
\[
| \hat \bB_i - \bB_i|  \leq C | \bB_i^{-1} - \hat \bB_i^{-1}|,
\]
and since the right hand side converges uniformly in $i$, \eqref{e:B} follows.

The result \eqref{e:ab} follows by combining \eqref{e:a} and \eqref{e:B}.  

We now turn to the final claim, \eqref{e:sig}.  By the same arguments above, we have that
\[
\sup_i | \hat \Sigma_i - \Sigma_i| = o_P(1).
\]
We now apply the same arguments as for \eqref{e:B}.  What isn't immediately obvious is that one can construct a closed convex set of positive definite matrices that contains all of the $\Sigma_i$.  This turns out to be possible as long as $\sigma^2_\vep > 0$ and $\sigma^2_\delta>0$, which we show in Section \ref{s:proj}.
\end{proof}

\section{Projection Calculation for $\Sigma_i$}\label{s:proj}
Here we show that one can construct a closed convex set within the cone of positive definite matrices that includes all of the $\Sigma_i$ and for any $N$.  Let $\mbK$ denote the RKHS generated from the covariance function of $C_X(t,s)$.  Now consider an arbitrary collection of elements of $\mbK$, denoted as $h_1,\dots, h_K$.  Consider the matrix
\[
\bS \bK \bS^\top
\]
where
\[
\bS[i,j] =  \langle v_i, h_j \rangle_{\mbK} 
\]
and
\[
\bK^{-1}[i,j] = \langle h_i, h_j \rangle_{\mbK}.
\]
Notice that within the RKHS, by the reproducing property, point-wise evaluation is given by $g(t) = \langle C_t, g \rangle_{\mbK}$, thus this setup is a bit more general than strictly needed.  
Now consider the operator
\[
P(x) = \sum_{k=1}^m h_k \sum_{s=1}^m K_{ks} \langle h_s, x\rangle_\mbK.
\]
Then, by direct verification, one can see that
\[
(\bS \bK \bS^\top)[i,j] = \langle v_i, P(v_j) \rangle_\mbK.
\]
Now one can easily verify that $P$ is a projection.  So we have that
\[
(\bS \bK \bS^\top)[i,j] = \langle P v_i, Pv_j \rangle_{\mbK}.  
\]
Now consider that 
\[
\langle v_i, v_j \rangle_\mbK = \langle P v_i + Q v_i , P v_j +  Q v_j \rangle_\mbK
= \langle P v_i ,   P v_j \rangle_\mbK + \langle Q v_i ,   Q v_j \rangle_\mbK.
\]
So, we have that the matrix of inner products in $\mbK$ can be expressed as the sum of two positive definite matrices.  

Returning to our original problem, let $h_1 = C_{XY}$ and $h_{j+1} =  C_{t_j}$ for $j=1,\dots,m$.  Briefly assume that $\sigma^2_\varepsilon = \sigma^2_\delta =0$, we will address these two quantities at the end.  Now, consider the scaled functions $\tilde v_j = \lambda_j v_j = C_X v_j$.  Then we have that
\[
\lambda_i 1_{ij} = \lambda_i \langle v_i, v_j \rangle = \langle \tilde v_i , \tilde v_j \rangle_{\mbK}. 
\]  
Now we need to verify that $\bA_i$ and $\bB_i^{-1}$ are of the form given for $\bS$ and $\bK$ respectively.  Notice that
\[
\langle \tilde v_j, C_{t} \rangle_\mbK = \tilde v_j (t) = \lambda_j v_j(t),
\]  
and that
\[
\langle \tilde v_j, C_{XY} \rangle_{\mbK} =\langle C_X v_j, C_{XY} \rangle_{\mbK} = \langle v_j, C_{XY} \rangle,
\]
thus the $\bA_i$ as desired.    Now by the reproducing property we have
\[
\langle C_t, C_s \rangle_{\mbK} = C(t,s) \qquad \text{and} \qquad
\langle C_{XY}, C_t \rangle_{\mbK} = C_{XY}(t).
\]
Lastly, recall that $C_{XY} = C_{X}\beta$ and $\sigma_Y^2 = \langle C_X \beta, \beta \rangle$, thus 
\[
\langle C_{XY}, C_{XY} \rangle_{\mbK} = 
\langle C_{X} \beta, C_X \beta_{\mbK} = \langle C_{X} \beta, \beta\rangle,
\]
as desired.  Thus, if $\sigma_\epsilon^2 = 0$ and $\sigma_{\delta}^2=0$, then this is a direct projection calculation and  we have that
\[
\Sigma_i = diag\{\lambda_1, \dots, \lambda_p\} - \bA_i \bB_i^{-1} \bA_i^T
\]
is positive definite since it can be expressed as a matrix of the pairwise inner products of $Q v_i$.  Finally, if we assume that $\sigma_\epsilon$ and $\sigma_{\delta}$ are not zero, then this has the effect of adding directly the diagonal of $\bB$.  Since $\bB^{-1}$ is positive definite, this means that as $\sigma_\epsilon$ and $\sigma_{\delta}$ increase, $\bB$ is strictly decreasing (in the sense of positive definite matrices).  Thus we can construct a closed convex set of positive definite matrices that contains all of the $\Sigma_i$ for any $i$ and all $N$.


\section{Additional Simulation Results}\label{sec:apptables}

\begin{table}[!h]
\centering
\scalebox{1}{
\begin{tabular}{lllllllllllll}
  \hline
   N & \multicolumn{4}{c}{w = 0} & \multicolumn{4}{c}{w = 5} & \multicolumn{4}{c}{w = 10} \\ \cmidrule(lr){2-5}\cmidrule(lr){6-9}\cmidrule(lr){10-13}
  & MeC & MuC & MeU & MuU & MeC & MuC & MeU & MuU & MeC & MuC & MeU & MuU \\
 \hline
100 & 1.44 & 0.05 & 1.44 & 0.05 & 5.53 & 3.34 & 6.57 & 7.76 & 14.08 & 13.19 & 22.90 & 30.88 \\ 
  200 & 0.74 & 0.02 & 0.74 & 0.02 & 5.26 & 3.31 & 5.02 & 7.73 & 13.96 & 13.17 & 17.99 & 30.95 \\ 
  400 & 0.32 & 0.01 & 0.32 & 0.01 & 5.12 & 3.30 & 4.13 & 7.73 & 13.90 & 13.16 & 15.42 & 30.90 \\ 
  800 & 0.18 & 0.01 & 0.18 & 0.01 & 5.03 & 3.29 & 3.69 & 7.70 & 13.87 & 13.15 & 14.27 & 30.76 \\ 
   \hline
\end{tabular}
}
\caption{MISE as N increases, using true parameters, with J = 3.} 
\end{table}

\begin{table}[!h]
\centering
\scalebox{1}{
\begin{tabular}{lllllllllllll}
  \hline
   m & \multicolumn{4}{c}{w = 0} & \multicolumn{4}{c}{w = 5} & \multicolumn{4}{c}{w = 10} \\ \cmidrule(lr){2-5}\cmidrule(lr){6-9}\cmidrule(lr){10-13}
  & MeC & MuC & MeU & MuU & MeC & MuC & MeU & MuU & MeC & MuC & MeU & MuU \\
 \hline
2 & 0.75 & 0.02 & 0.75 & 0.02 & 5.26 & 3.31 & 5.04 & 7.73 & 13.96 & 13.17 & 18.02 & 30.95 \\ 
  5 & 0.31 & 0.04 & 0.31 & 0.04 & 4.26 & 3.32 & 3.77 & 4.88 & 13.76 & 13.17 & 14.25 & 19.33 \\ 
  10 & 0.18 & 0.05 & 0.18 & 0.05 & 3.74 & 3.33 & 3.53 & 3.84 & 13.58 & 13.18 & 13.61 & 15.18 \\ 
  20 & 0.12 & 0.06 & 0.12 & 0.06 & 3.51 & 3.35 & 3.45 & 3.50 & 13.42 & 13.20 & 13.41 & 13.81 \\ 
   \hline
\end{tabular}
}
\caption{MISE as m increases, using true parameters, with J = 3.} 
\end{table}

\begin{table}[!h]
\centering
\scalebox{1}{
\begin{tabular}{lllllllllllll}
  \hline
   N & \multicolumn{4}{c}{w = 0} & \multicolumn{4}{c}{w = 5} & \multicolumn{4}{c}{w = 10} \\ \cmidrule(lr){2-5}\cmidrule(lr){6-9}\cmidrule(lr){10-13}
  & MeC & MuC & MeU & MuU & MeC & MuC & MeU & MuU & MeC & MuC & MeU & MuU \\
 \hline
100 & 392.04 & 0.40 & 392.04 & 0.40 & 245.08 & 0.44 & 946.97 & 8.52 & 73.18 & 0.51 & 2492.00 & 33.28 \\ 
  200 & 177.30 & 0.19 & 177.30 & 0.19 & 198.40 & 0.21 & 431.76 & 8.07 & 59.15 & 0.34 & 1127.66 & 31.70 \\ 
  400 & 83.97 & 0.09 & 83.97 & 0.09 & 167.24 & 0.13 & 207.07 & 7.84 & 51.47 & 0.26 & 599.76 & 31.07 \\ 
  800 & 45.39 & 0.04 & 45.39 & 0.04 & 153.83 & 0.09 & 108.69 & 7.61 & 47.96 & 0.22 & 314.43 & 30.39 \\ 
   \hline
\end{tabular}
}
\caption{MISE as N increases, using true parameters, with J = 5.} 
\end{table}

\begin{table}[!h]
\centering
\scalebox{1}{
\begin{tabular}{lllllllllllll}
  \hline
   m & \multicolumn{4}{c}{w = 0} & \multicolumn{4}{c}{w = 5} & \multicolumn{4}{c}{w = 10} \\ \cmidrule(lr){2-5}\cmidrule(lr){6-9}\cmidrule(lr){10-13}
  & MeC & MuC & MeU & MuU & MeC & MuC & MeU & MuU & MeC & MuC & MeU & MuU \\
 \hline
2 & 177.30 & 0.19 & 177.30 & 0.19 & 198.73 & 0.21 & 431.76 & 8.07 & 59.15 & 0.34 & 1127.66 & 31.70 \\ 
  5 & 54.93 & 0.24 & 54.93 & 0.24 & 96.32 & 0.27 & 100.12 & 4.60 & 43.51 & 0.37 & 264.40 & 17.75 \\ 
  10 & 27.38 & 0.32 & 27.38 & 0.32 & 47.51 & 0.32 & 32.93 & 3.03 & 31.71 & 0.42 & 62.47 & 11.08 \\ 
  20 & 11.37 & 0.44 & 11.37 & 0.44 & 20.66 & 0.43 & 13.52 & 2.34 & 19.97 & 0.50 & 21.54 & 7.65 \\ 
   \hline
\end{tabular}
}
\caption{MISE as m increases, using true parameters, with J = 5.} 
\end{table}

\end{document}